\documentclass[aps,prd,nofootinbib,amsmath,superscriptaddress]{revtex4}

\usepackage{bm}
\usepackage{graphicx}
\usepackage{xcolor}
\newcommand{\de}{{\rm d}}
\newcommand{\be}{\begin{equation}}
\newcommand{\ee}{\end{equation}}
\newcommand{\bea}{\begin{eqnarray}}
\newcommand{\eea}{\end{eqnarray}}

\newcommand{\bfk}{\mbox{\boldmath $k$}}
\def\bkt{\bfk_\perp}
\def\kt{k_\perp}
\newcommand{\bfp}{\mbox{\boldmath $p$}}

\newcommand{\bfq}{\mbox{\boldmath $q$}}
\newcommand{\bfP}{\mbox{\boldmath $P$}}

\def\pp{p_\perp}
\newcommand{\pup}{p^\uparrow}

\def\avk{\langle k_\perp ^2\rangle}
\def\avp{\langle p_\perp ^2\rangle}

\def\avPT{\langle P_T^2\rangle}

\def\C{_{_C}}

\begin{document}

\title{Role of transverse momentum dependence of unpolarised parton distribution and\\
fragmentation functions in the analysis of azimuthal spin asymmetries}

\author{M.~Anselmino}
\email{mauro.anselmino@to.infn.it}
\affiliation{Dipartimento di Fisica, Universit\`a di Torino,
             Via P.~Giuria 1, I-10125 Torino, Italy}
\affiliation{INFN, Sezione di Torino, Via P.~Giuria 1, I-10125 Torino, Italy}
\author{M.~Boglione}
\email{elena.boglione@to.infn.it}
\affiliation{Dipartimento di Fisica, Universit\`a di Torino,
             Via P.~Giuria 1, I-10125 Torino, Italy}
\affiliation{INFN, Sezione di Torino, Via P.~Giuria 1, I-10125 Torino, Italy}
\author{U.~D'Alesio}
\email{umberto.dalesio@ca.infn.it}
\affiliation{Dipartimento di Fisica, Universit\`a di Cagliari,
             Cittadella Universitaria, I-09042 Monserrato (CA), Italy}
\affiliation{INFN, Sezione di Cagliari,
             Cittadella Universitaria, I-09042 Monserrato (CA), Italy}
\author{F.~Murgia}
\email{francesco.murgia@ca.infn.it}
\affiliation{INFN, Sezione di Cagliari,
             Cittadella Universitaria, I-09042 Monserrato (CA), Italy}
\author{A.~Prokudin}
\email{prokudin@jlab.org}
\affiliation{Science Division, Penn State University Berks, Reading, Pennsylvania 19610, USA}
\affiliation{Theory Center, Jefferson Lab, 12000 Jefferson Avenue, Newport News, VA 23606, USA}
\date{\today}

\begin{abstract}
Information on the Sivers distribution and the Collins fragmentation functions
and their transverse momentum dependence is mainly based on fitting single spin
asymmetry data from semi-inclusive deep inelastic scattering (SIDIS). Independent
information, respectively on the Sivers distribution and the Collins
fragmentation, can be obtained from Drell-Yan and $e^+e^-$ annihilation
processes. In the SIDIS case, the transverse momentum of the
final observed hadron, which is the quantity measured, is generated both by
the average transverse momentum in the distribution and in the fragmentation
functions. As a consequence, these are strongly correlated and a separate
extraction is made difficult. In this paper we investigate, in a simple
kinematical Gaussian configuration, this correlation, its role on the
transverse single spin asymmetries in SIDIS and the consequences for predictions
of the Sivers asymmetry in Drell-Yan processes and for the Collins asymmetry
in $e^+e^-$ annihilation. We find that, in some cases, these effects can be relevant
and must be carefully taken into account.
\end{abstract}


\maketitle

\section{\label{sec:intro} Introduction}

Transverse Momentum Dependent Parton Distribution and Fragmentation Functions
(respectively TMD PDFs and TMD FFs, collectively denoted as TMDs) are important
tools for investigating the nucleon and its three-dimensional (3D) structure. Among
them, the Sivers function~\cite{Sivers:1989cc,Sivers:1990fh} describes the
asymmetry in the azimuthal distribution of unpolarised quarks and gluons around
the direction of motion of a high-energy transversely polarised parent hadron.
Similarly, the Collins fragmentation function~\cite{Collins:1992kk}
gives the azimuthal distribution of unpolarised hadrons around the
direction of motion of a transversely polarised fragmenting quark. The former
is related to the orbital motion of partons inside a nucleon, while the latter
describes fundamental properties of the hadronisation process.

Azimuthal and transverse single spin asymmetries (SSAs) in inclusive and semi-inclusive
hadron production are the fundamental source of information on these non
perturbative functions. The Sivers and the Collins effects indeed play a crucial
role in describing, within the so-called TMD factorisation approach, many of the
transverse and azimuthal asymmetries experimentally observed in semi-inclusive deep
inelastic scattering (SIDIS) and in $e^+e^-$ annihilations. The Sivers asymmetry is
also crucial for understanding the single spin asymmetries in polarised Drell-Yan
processes, although experimental information in this case is still scarce.
The transverse momentum dependence of the unpolarised TMDs is related
to the $P_T$ distribution of hadrons produced in unpolarised SIDIS processes.

The first phase in the extraction of the TMDs from data can now be considered as
complete. It has shown that the Sivers and Collins effects are indeed significant~\cite{Airapetian:2009ae, Adolph:2012sp,Allada:2013nsw,Airapetian:2010ds,
Adolph:2012sn,Adolph:2014zba,Abe:2005zx,Seidl:2008xc,Aubert:2015hha,
TheBABAR:2013yha,Ablikim:2015pta}, and information on the Sivers and Collins functions is now available~\cite{Vogelsang:2005cs,Collins:2005ie,Anselmino:2008sga,Anselmino:2008jk,Bacchetta:2011gx,
Anselmino:2013vqa,Anselmino:2015sxa,Anselmino:2015fty,Anselmino:2016uie}.
Using the Collins effect, also the extraction of the quark transversity
distribution has been possible~\cite{Anselmino:2007fs}. In this phase a very
simple parameterisation of the unknown functions has been adopted, with
factorised dependences on the different variables and a simple (and analytically
integrable) Gaussian dependence on the transverse momenta. Thanks to important
theoretical progress a second phase has now started in which the QCD TMD
evolution can be taken into account and a global fit of data from different
processes can be attempted~\cite{Bacchetta:2013pqa,Signori:2013mda,Anselmino:2013lza,Echevarria:2014xaa,%
DAlesio:2014mrz,Kang:2015msa,Bacchetta:2017gcc,Scimemi:2017etj}.
More refined and realistic parameterisations of the TMDs
can be explored, leading to their more precise determination.

Before entering this phase, some considerations about the procedure
of extraction of TMDs from data and the combined analysis of different processes
are necessary. This concerns the way in which TMDs build up the measured quantities
and the fact that often two of them are coupled into a unique observable;
thus, disentangling information
on a single TMD is not always straightforward and could lead to uncertainties which have
to be taken into account. Here, we do this in the simple approach of the first phase mentioned above,
which much simplifies and exemplifies the issues to be discussed, without
spoiling their general features.

We start by noticing that most of the available information on spin
asymmetries - related to the Sivers and Collins functions - and on unpolarised
TMDs, is obtained from SIDIS processes data. In this case, however, the
transverse momentum of the final observed hadron, $\bm{P}_{T}$, originates
both from the transverse motion of the initial struck quark inside the nucleon,
$\bfk_\perp$, and the transverse momentum of the final hadron with respect to
the fragmenting quark, $\bfp_\perp$. At leading order in a $k_\perp/Q$ power expansion,
where $Q$ is the hard scale for the process considered, one has
\be
\bm{P}_{T} = \bm{p}_\perp +z\bm{k}_\perp \>,\label{PhT}
\ee
where $z$ is the light-cone momentum fraction of the hadron in the quark fragmentation process.
As a consequence, in SIDIS, the transverse momentum dependences in the initial
quark TMD-PDFs and in the fragmenting quark TMD-FFs are strongly correlated,
as it has been already pointed out~\cite{Signori:2013mda,Anselmino:2013lza,Bacchetta:2017gcc}.

This dependence is usually parameterised by a
Gaussian function, in which the main parameters are the widths $\avk$ (for TMD-PDF)
and $\avp$ (for TMD-FF). Because of the relation given in Eq.~(\ref{PhT}), it is
possible to obtain good fits of SIDIS data, with comparable $\chi^2_{\rm dof}$,
corresponding to different pairs of values for $\avk$ and $\avp$.
However, the parameters of these comparable fits may lead to rather different
consequences when used to get estimates for asymmetries in processes in which
only TMD-PDFs, like Drell-Yan processes, or only TMD-FFs, like
two-hadron production in $e^+e^-$ annihilations, are involved.

In this paper we investigate this issue in more details. To this end, we consider,
in the TMD factorisation approach of the first phase, the Sivers and Collins
transverse single spin asymmetries, together with the corresponding unpolarised
cross sections, in SIDIS and Drell-Yan processes and in $e^+e^-$ annihilations.
The plan of the paper is the following: in section~\ref{sec:gen} we will present
the general expressions for the unpolarised cross sections and the single spin
asymmetries of interest for our study, referring to the original literature
for their derivation. In section~\ref{sec:siv} we shall consider the
study of the Sivers asymmetry in SIDIS and Drell-Yan processes, while
in section~\ref{sec:col} we will discuss the Collins asymmetries in SIDIS and
$e^+e^-$ annihilations. Finally, in section~\ref{sec:conclusions} we will
summarise our main results and their possible consequences for future
studies of azimuthal and single spin asymmetries in Drell-Yan processes and $e^+e^-$
annihilations.

\section{\label{sec:gen} General results in the TMD approach}

In this section we recall the formalism which we shall need for our
discussion about the extraction of the transverse momentum dependence of
the TMDs. In semi-inclusive DIS, TMD factorisation theorems \cite{Collins:2011zzd,Aybat:2011ge,%
GarciaEchevarria:2011rb,Echevarria:2012js,Echevarria:2014rua,Rogers:2015sqa,Gonzalez-Hernandez:2018ipj}
relate the transverse momentum of the produced hadron to the intrinsic transverse momenta
of the parton inside the target nucleon and in the quark hadronisation process.
Such factorisation theorems, and the analogous ones for Drell-Yan processes
and $e^+e^-$ annihilations, are controllable approximations that allow one to relate the observed cross sections to convolutions of TMDs. Even though generic constraints on the functional form of the non perturbative functions are given by the theorems themselves, the phenomenological analysis of the experimental data is needed to determine the functional shape of the TMDs.

We present the explicit expressions of the measured quantities within the TMD factorisation approach at parton model level and with Gaussian parameterisations for the TMDs; references to the original papers are given.
It is however convenient to remind here the parameterisations adopted for the
relevant TMDs. A parton inside a nucleon with momentum $\bfP$ has a momentum
$\bfp = x\,\bfP + \bkt$, while a hadron produced in the fragmentation of a quark with
momentum $\bfp_q$ has a momentum $\bfP_h = z \, \bfp_q + \bfp_\perp$.
Notice that at leading order in a $k_\perp/Q$ power expansion
longitudinal and light-cone momentum fractions coincide,
neglecting quark and final hadron masses.

The unpolarised TMD-PDFs and TMD-FFs are respectively chosen
as~\cite{Anselmino:2011ch}:
\be
f_{q/p} (x,\kt)= f_{q/p} (x)\,\frac{e^{-\kt^2/\avk}}{\pi\avk}
\quad\quad\quad
D_{h/q}(z,\pp)=D_{h/q}(z)\,\frac{e^{-\pp^2/\avp}}{\pi\avp}\,,
\label{unp-TMD}
\ee
while the Sivers function is written as
\be
\Delta^N f_{q/\pup} (x,\kt) = \Delta^N f_{q/\pup}(x)\;
\sqrt{2e}\,\frac{\kt}{M_S} \; e^{-\kt^2/M^2_S}\,
\frac{e^{-\kt^2/\avk}}{\pi\avk}
\equiv \Delta^N f_{q/\pup}(x) \; \sqrt{2e}\,\frac{\kt}{M_S} \;
\frac{e^{-\kt^2/\avk_S}}{\pi\avk}
\label{Siv-dist}
\ee
and the Collins function as
\be
\Delta^N  D_{h/q^\uparrow}(z,\pp) = \Delta^N  D_{h/q^\uparrow}(z)\;
\sqrt{2e}\,\frac{\pp}{M_C} \; e^{-\pp^2/M_C^2}\,
\frac{e^{-\pp^2/\avp}}{\pi\avp}
\equiv \Delta^N  D_{h/q^\uparrow}(z)\;
\sqrt{2e}\,\frac{\pp}{M_C} \;
\frac{e^{-\pp^2/\avp \C}}{\pi\avp}\,,\label{Coll-frag}
\ee
where we have defined
\be
\avk_S = \frac{\avk \, M^2_S}{\avk +  M^2_S}
\quad\quad\quad
\avp_C=\frac{\avp \, M_C^2}{\avp + M_C^2}\,\cdot \label{S-C-2}
\ee
These functional shapes are particularly suitable
in order to directly impose the known positivity bounds on the Sivers and Collins functions.
Notice that the factorised transverse momentum dependences have a Gaussian
shape with a width which is constant and flavour independent.

\subsection{\label{sec:gen:sidis-siv} The Sivers SSA in the SIDIS process
$\ell p^\uparrow\to \ell^\prime h\,X$}

Following Ref.~\cite{Anselmino:2011ch} (see also Ref.~\cite{Bacchetta:2006tn}), where all details can be found,
the differential cross section for the semi-inclusive production of a hadron $h$,
in the current fragmentation region, from the collision of an unpolarised
lepton beam off a transversely polarised target can be written, in the deeply
inelastic regime, as (see Eq.~(79) of Ref.~\cite{Anselmino:2011ch}):
\bea
\dfrac{\de\sigma^{\ell p(S_T)\to \ell^\prime h\,X}}
{\de x_{B}\, \de Q^2 \, \de z_h \, \de ^2\bm{P}_{T} \, \de \phi_S} &=&
\dfrac{2\alpha^2}{Q^4}\,\Bigl\{ \frac{1+(1-y)^2}{2}\,F_{UU} + \dots
\label{eq:dsig-sidis} \\
&+& \Bigl[ \frac{1+(1-y)^2}{2}\,\sin(\phi_h-\phi_S)\,
F_{UT}^{\sin(\phi_h-\phi_S)} +
(1-y)\,\sin(\phi_h+\phi_S)\,F_{UT}^{\sin(\phi_h+\phi_S)} + \dots\,\Bigr]\Bigr\}\,.
\nonumber
\eea

We have considered the case of a transversely polarised target ($S_T = 1,
S_L = 0$) and unpolarised beam ($P_z^\ell = 0$); we have omitted terms which
are not related to the Sivers or Collins asymmetries.
$x_B$, $y$, $z_h$ and $Q$ are the usual SIDIS variables.
Notice that, at order $k_\perp/Q$, $x_B = x$ and $z_h = z$.
$P_T$ is the magnitude of the hadron transverse momentum in the
$\gamma$*-nucleon c.m.~frame; $\phi_h$ and $\phi_S$ are respectively the azimuthal
angle of the observed hadron and of the target polarisation vector w.r.t.~the
leptonic plane. The subscript $UT$ in the structure functions $F$
reminds that we are considering the case of an unpolarised lepton beam and a
transversely polarised nucleon target ($UU$ refers to the unpolarised situation).

In the SIDIS case, the asymmetries are expressed through their azimuthal
moments,
\be
A_{UT}^{W(\phi_h,\phi_S)} = 2 \,
\frac{\int\,d\phi_h d\phi_S\,\left[ d\sigma^\uparrow -
d\sigma^\downarrow\right]\,W(\phi_h,\phi_S)}
{\int\,d\phi_h d\phi_S\,\left[ d\sigma^\uparrow + d\sigma^\downarrow\right]}\,,
\label{eq:azi-mom}
\ee
where $W(\phi_h,\phi_S)$ is the appropriate azimuthal weight function
required in order to isolate the specific contribution of interest and
$d\sigma^{\uparrow,\downarrow}$ is the differential cross section of
Eq.~(\ref{eq:dsig-sidis}) with $S_T = \> \uparrow,\downarrow$ denoting,
respectively, a transverse polarisation with azimuthal angle $\phi_S$ and
$\phi_S + \pi$. Then we simply have
\bea
d\sigma^\uparrow - d\sigma^\downarrow &=&
\frac{2\alpha^2}{Q^4}\,\Bigl\{ [1+(1-y)^2]\,\sin(\phi_h -
\phi_S)\,F_{UT}^{\sin(\phi_h-\phi_S)} +
2(1-y)\,\sin(\phi_h + \phi_S)\,F_{UT}^{\sin(\phi_h+\phi_S)} + \dots \Bigr\}\,,
\label{eq:dsig-updown1}\\
d\sigma^\uparrow + d\sigma^\downarrow &=&
\frac{2\alpha^2}{Q^4}\,\Bigl\{ [1+(1-y)^2]\,F_{UU} + \dots \Bigr\}\,.
\label{eq:dsig-updown2}
\eea
The Sivers asymmetry is related to the $\sin(\phi_h - \phi_S)$ modulation
and from Eqs.~(\ref{eq:azi-mom})-(\ref{eq:dsig-updown2}) we find
\be
A_{UT}^{\sin(\phi_h - \phi_S)} =
\dfrac{
\,F_{UT}^{\sin(\phi_h-\phi_S)}\,}
{F_{UU}}\,\cdot
\label{eq:AUT-siv}
\ee

{}From Eqs.~(115) and (123) of Ref.~\cite{Anselmino:2011ch} (remember
that $x_B = x$ and $z_h = z$) we see that
\bea
F_{UU} &=& \sum_q\,e_q^2\,f_{q/p}(x)\,D_{h/q}(z)\,
\dfrac{e^{-P_T^2/\langle P_T^2\rangle}}{\pi\langle P_T^2\rangle}
\label{eq:F-uu}\\
F_{UT}^{\sin(\phi_h - \phi_S)} &=& \sum_q\,e_q^2\,\Delta^N f_{q/p^\uparrow}(x)
\,D_{h/q}(z)
\, \sqrt{\frac{e}{2}}\,\frac{P_T}{M_S}\,\frac{z\,\langle k_\perp^2\rangle_S^2}
{\langle k_\perp^2\rangle}\,
\dfrac{e^{-P_T^2/\langle P_T^2\rangle_S}}{\pi\langle P_T^2\rangle_S^2}\,,
\label{eq:F-ut-siv}
\eea
where (see Eq.~(131) of Ref.~\cite{Anselmino:2011ch}):
\be
\avPT = \avp + z^2\,\avk
\quad\quad\quad
\avPT_S = \avp + z^2 \avk_S \,,
\label{eq:PT-S}
\ee
with $\avk_S$ as in Eq.~(\ref{S-C-2}).

These relations, valid at first order in a $k_\perp/Q$ power expansion,
show explicitly the strong correlation, in building the physical observables,
between the properties of the partonic transverse momentum distribution and those
of the partonic fragmentation. It is, in fact, the analysis of these entangled
effects which motivates our study. Notice that this correlation is also
modulated by the value of $z$.

{}From Eqs.~(\ref{eq:AUT-siv})-(\ref{eq:F-ut-siv}) we see that the Sivers azimuthal
asymmetry for SIDIS processes can be factorised as
\be
A_{UT}^{\sin(\phi_h-\phi_S)}(x,z,P_T) =
 A^S_{\rm DIS}(x,z)\,F^S_{\rm DIS}(z,P_T)\,,
\label{eq:A-UT-AF}
\ee
where
\bea
A^S_{\rm DIS}(x,z) &=& \dfrac{
\sum_q\,e_q^2\,\Delta^N f_{q/p^\uparrow}(x)\,D_{h/q}(z)}
{2\,
\sum_q\,e_q^2\,f_{q/p}(x)\,D_{h/q}(z)}
\label{eq:A-S-DIS}\\
F^S_{\rm DIS}(z,P_T) &=&
\dfrac{\sqrt{2e}\,\dfrac{P_T}{M_S}\,\dfrac{z\,\langle k_\perp^2\rangle_S^2\,
\exp[-P_T^2/\langle P_T^2\rangle_S]}{\pi\,\langle k_\perp^2\rangle\,
\langle P_T^2\rangle_S^2}}{\dfrac{\exp[-P_T^2/\langle P_T^2\rangle]}{\pi\,\langle
P_T^2\rangle}} \> \cdot
\label{eq:F-S-DIS}
\eea

If we now integrate separately the numerator and the denominator of
$F^S_{\rm DIS}$ over the modulus of the transverse momentum of the observed
hadron, $P_T\,\de P_T$, in the full $P_T$-range $[0,+\infty)$, and define the
dimensionless parameters
\be
\xi_1 = \frac{\langle p_\perp^2\rangle}{\langle k_\perp^2\rangle}
\quad\quad\quad
\rho_S = \frac{\avk_S}{\avk} =
\frac{M_S^2}{M_S^2 + \avk} \>,
\label{eq:rho_s}
\ee
we find the $P_T$-integrated Sivers asymmetry for SIDIS,
\be
A_{UT}^{\sin(\phi_h-\phi_S)}(x,z) =
 A^S_{\rm DIS}(x,z)\,{\cal F}^S_{\rm DIS}(z)\,,
\label{eq:A-UT-AF-int}
\ee
where
\be
{\cal F}^S_{\rm DIS}(z,\xi_1, \rho_S) = \sqrt{\frac{e\pi}{2}}\,\left[\,
\dfrac{\rho_S^3(1-\rho_S)}{\rho_S + \xi_1/z^2}\right]^{1/2}\,.
\label{eq:cal-F-S-DIS}
\ee
Notice that $0 < \rho_S < 1$. For $M_S^2 \ll \langle k_\perp^2\rangle$,
$\rho_S \to 0$; in this case, the $k_\perp$ dependent part of the Sivers function
is sharply peaked around zero and, at its maximum, almost equals the $k_\perp$ dependent component of the
unpolarised distribution.
On the other hand, for $M_S^2 \gg \langle k_\perp^2\rangle$, $\rho_S \to 1$; correspondingly,
the $k_\perp$ dependent part of the Sivers function
is peaked around $\sqrt{\langle k_\perp^2\rangle/2}$, where its value becomes smaller and smaller.
Both these borderline cases are not very relevant from the phenomenological point of view,
although for completeness we shall consider the full range of values for $\rho_S$.

A comment on the $P_T\,\de P_T$ integration, which applies as well to the
next subsections, is necessary. Such an integration can be performed
analytically and leads to very simple results, but it exceeds the range of
validity of the TMD factorisation, which holds up to transverse momenta of
the order of a few GeV only, such that $P_T/z \ll Q$. Above that, higher order QCD corrections become
dominant. However, because of the Gaussian dependences, the large $P_T$
values do not contribute significantly to the integrations, which are indeed
dominated by the region of validity of the TMD factorisation. Our fully $P_T$-integrated
expressions can be safely compared with data collected at small
$P_T$ values ($P_T$ up to 1-2 GeV).

\subsection{\label{sec:gen:sidis-coll} The Collins SSA in SIDIS processes}

The Collins effect generates a $\sin(\phi_h + \phi_S)$ modulation, and from
Eqs.~(\ref{eq:azi-mom}), (\ref{eq:dsig-updown1}) and (\ref{eq:dsig-updown2}), we find that the
azimuthal moment of the Collins asymmetry in SIDIS processes can be written as
\be
A_{UT}^{\sin(\phi_h + \phi_S)} =
\dfrac{2(1-y)}{ 1+(1-y)^2}\,\dfrac{\,F_{UT}^{\sin(\phi_h+\phi_S)}\,}{F_{UU}}\,,
\label{eq:AUT-coll}
\ee
where $F_{UU}$ is given by Eq.~(\ref{eq:F-uu}) and $F_{UT}^{\sin(\phi_h+\phi_S)}$
can be taken from Eq.~(127) of Ref.~\cite{Anselmino:2011ch} (noticing that the
parameter $M_h$ is here denoted as $M_C$)
:
\be
F_{UT}^{\sin(\phi_h + \phi_S)} =
\sum_q\,e_q^2\,h_1^q(x)\,\Delta^N D_{h/q^\uparrow}(z)
\, \sqrt{\frac{e}{2}}\,\frac{P_T}{M_C}\,\frac{\avp_C^2}{\avp}\,
\dfrac{e^{-P_T^2/\avPT_T}}{\pi\avPT_T^2}\,\cdot
\label{eq:F-ut-coll}
\ee
In this equation $h_1^q(x)$ is the $k_\perp$-integrated, collinear quark transversity
distribution, $\Delta^N D_{h/q^\uparrow}(z)$ is the $z$-dependent term in the
Collins fragmentation function (see Eq.~(\ref{Coll-frag})), $\avk_T$ is the
flavour-independent average square transverse momentum for the transversity distribution and
\be
\avPT_T = \avp_C + z^2\,\avk_T\,,
\label{eq:PT2-T}
\ee
with $\avp_C$ given in Eq.~(\ref{S-C-2}).

In complete analogy to the Sivers asymmetry, also in the Collins case we can
write
\be
A_{UT}^{\sin(\phi_h+\phi_S)}(x,y,z,P_T) = A^C_{\rm DIS}(x,y,z)
\,F^C_{\rm DIS}(z,P_T)\,,
\label{eq:A-UT-AF-coll}
\ee
where
\bea
A^C_{\rm DIS}(x,y,z) &=& \dfrac{1-y}{1+(1-y)^2} \,
\dfrac{\sum_q\,e_q^2\,h_1^q(x)\,\Delta^N D_{h/q^\uparrow}(z)}
{\sum_q\,e_q^2\,f_{q/p}(x)\,D_{h/q}(z)}
\label{eq:A-S-DIS-coll}\\
\nonumber\\
F^C_{\rm DIS}(z,P_T) &=&
\dfrac{\sqrt{2e}\,\dfrac{P_T}{M_C}\,\dfrac{\langle p_\perp^2\rangle_C^2\,
\exp[-P_T^2/\langle P_T^2\rangle_T]}{\pi\,\langle p_\perp^2\rangle\,
\langle P_T^2\rangle_T^2}}{\dfrac{\exp[-P_T^2/\langle P_T^2\rangle]}{\pi\,\langle P_T^2\rangle}}\,\cdot
\label{eq:F-S-DIS-coll}
\eea

Once more, integrating separately the numerator and denominator of $F^C_{\rm DIS}$
over $P_T \, {\rm d}P_T$ in the full range $[0,+\infty)$, and defining the
dimensionless parameters
\be
\xi_T = \frac{\avk_T}{\avk} \quad\quad\quad
\rho_C = \frac{\avp_C}{\avp} = \dfrac{M_C^2}{M_C^2 + \avp} \>,
\ee
we can write the $P_T$-integrated Collins asymmetry for SIDIS as:
\be
A_{UT}^{\sin(\phi_h+\phi_S)}(x,y,z) = A^C_{\rm DIS}(x,y,z)
\,{\cal F}^C_{\rm DIS}(z)\,,
\label{eq:A-UT-AF-coll-int}
\ee
with
\be
{\cal F}^C_{\rm DIS}(z,\rho_C,\xi_1/\xi_T) = \sqrt{\frac{e\pi}{2}}\,\left[\,
\dfrac{\rho_C^3(1-\rho_C)}{\rho_C + z^2(\xi_{T}/\xi_1)}\right]^{1/2}\,.
\label{eq:cal-F-C-DIS}
\ee
Notice the similarity with ${\cal F}^S_{\rm DIS}$, Eq.~(\ref{eq:cal-F-S-DIS}).

\subsection{\label{sec:gen:dy} The Sivers SSA in Drell-Yan processes,
$h_1^\uparrow h_2\to \ell^+\ell^-\,X$ }

Similarly to the SIDIS case, the Sivers asymmetry to be measured in DY
processes is (see Ref.~\cite{Anselmino:2009st} for all details):
\be
A_N^{\sin(\phi_{\gamma} - \phi_S)} \equiv A_N^{\rm DY}(y,M,q_T)=
2\,\frac{\int d\phi_{\gamma} \>
[d\sigma^{\uparrow} - d\sigma^{\downarrow}] \>
\sin(\phi_{\gamma}-\phi_S)}
{\int d\phi_{\gamma} \>
[d\sigma^{\uparrow} + d\sigma^{\downarrow}]} \>, \label{ANW}
\ee
where $d\sigma^{\uparrow,\downarrow}$ stands here for the cross section
\be
\frac{d^4\sigma^{h_1^{\uparrow,\downarrow} h_2\to \ell^+\ell^-\,X}}
{dy \, dM^2 \, d^2\bfq_T} \>,
\label{x-sect}
\ee
with $y$, $M$, and $\bfq_T$ being respectively the rapidity, the invariant mass,
and the transverse momentum of the final leptonic pair, while $\phi_\gamma$ and
$\phi_S$ are respectively the azimuthal angle of the virtual boson and of the transverse polarisation of the initial hadron in the c.m. frame of the two
colliding hadrons.

We limit our discussion to the energy regime $M\ll M_{W,Z}$, where
electromagnetic contributions dominate, neglecting weak interaction terms.
Following Ref.~\cite{Anselmino:2009st}, with the parameterisation of the TMDs
as in Eqs.~(\ref{unp-TMD}) and (\ref{Siv-dist}), the numerator and the denominator
of the SSA $A_N^{\rm DY}$ read:
\be
{\rm Num}[A_N^{\rm DY}] = \frac{4\pi\alpha^2}{9M^2 s}\,
\sum_q e_q^2\,\Delta^N f_{q/h_1^\uparrow}(x_1) f_{\bar q/h_2}(x_2)
\, \sqrt{2 e}\,\frac{q_T}{M_S}\,\frac{\langle k_\perp^2 \rangle_S^2\,
\exp[-q_T^2/(\langle k_\perp^2 \rangle_S + \langle k_{\perp 2}^2 \rangle)]}
{\pi\langle k_{\perp 1}^2 \rangle [\langle k_\perp^2 \rangle_S +
\langle k_{\perp 2}^2 \rangle]^2}\,
\label{eq:n-an-dy}
\ee
\be
{\rm Den}[A_N^{\rm DY}] = \frac{4\pi\alpha^2}{9M^2 s}\,2\,
\sum_q e_q^2\,f_{q/h_1}(x_1) f_{\bar q/h_2}(x_2) \, \frac{
\exp[-q_T^2/(\langle k_{\perp 1}^2 \rangle + \langle k_{\perp 2}^2 \rangle)]}
{\pi [\langle k_{\perp 1}^2 \rangle + \langle k_{\perp 2}^2 \rangle]}\,\cdot
\label{eq:d-an-dy}
\ee

Here $x_1$ and $x_2$ are, as usual, the light-cone momentum fractions of the
active quark and antiquark annihilating into the final lepton pair;
$\langle k_{\perp 1}^2\rangle$ and $\langle k_{\perp 2}^2\rangle$ are
the average square transverse momenta of the unpolarised quarks,
or antiquarks, inside the unpolarised initial hadrons. They are taken to be
flavour and $x_{1,2}$-independent. In general, they can
be different for different hadrons, like, for example, in the pion-proton DY processes measured at
COMPASS. At leading order in a $k_\perp/M$ power expansion, as it is well known,
one has
\be
x_1 = \frac{M}{\sqrt{s}}\,\,e^{y}\qquad\qquad x_2 = \frac{M}{\sqrt{s}}\,\,e^{-y}\,.
\label{eq:x12-def}
\ee

Again, the Sivers SSA $A_N^{\rm DY}$ factorises into two terms, one
$(x_1,x_2)$-dependent, and one $q_T$-dependent
\be
A_N^{\rm DY}(y,M,q_T) = A^S_{\rm DY}(x_1,x_2)\,F^S_{\rm DY}(q_T)\,,
\label{eq:A-DY-AF}
\ee
where
\bea
A^S_{\rm DY}(x_1,x_2) &\equiv& A^S_{\rm DY}(y,M) =
\frac{\sum_q e_q^2\,\Delta^N f_{q/h_1^\uparrow}(x_1) f_{\bar q/h_2}(x_2)}
{2\sum_q e_q^2\,f_{q/h_1}(x_1) f_{\bar q/h_2}(x_2)}
\label{eq:S-dy}\,,
\\
\nonumber\\
\nonumber\\
F^S_{\rm DY}(q_T) &=&
\dfrac{\sqrt{2e}\,\dfrac{q_T}{M_S}\,\dfrac{\langle k_\perp^2\rangle_S^2\,
\exp[-q_T^2/(\langle k_\perp^2\rangle_S + \langle k_{\perp 2}^2\rangle)]}{
\pi\,\langle k_{\perp 1}^2\rangle\,[\langle k_\perp^2\rangle_S +
\langle k_{\perp 2}^2\rangle)]^2}}
{\dfrac{\exp[-q_T^2/(\langle k_{\perp 1}^2\rangle + \langle k_{\perp 2}^2\rangle)]}
{\pi\,[\langle k_{\perp 1}^2\rangle + \langle k_{\perp 2}^2\rangle)]}}\,.
\label{eq:AF-dy-def}
\eea

By integrating separately the numerator and denominator of $F^S_{\rm DY}$
over $q_T \de q_T$ in the full range $[0,+\infty)$ and defining the dimensionless
parameter:
\be
\xi_{21} = \frac{\langle k_{\perp 2}^2\rangle}{\langle k_{\perp 1}^2\rangle}
\quad\quad\quad
\rho_S = \frac{\avk_S}{\langle k_{\perp 1}^2\rangle} =
\frac{M_S^2}{M_S^2 + \langle k_{\perp 1}^2\rangle}
\label{eq:xi-def}
\end{equation}
we get the $P_T$-integrated Sivers asymmetry for DY as
\be
A_N^{\rm DY}(y,M) = A^S_{\rm DY}(x_1,x_2)\,{\cal F}^S_{\rm DY}\,,
\label{eq:A-DY-AF-int}
\ee
with the simple expression
\be
{\cal F}^S_{\rm DY}(\rho_S,\xi_{21}) = \sqrt{\frac{e\pi}{2}}\,\left[\,
\dfrac{\rho_S^3(1-\rho_S)}{\rho_S+\xi_{21}}\right]^{1/2}\,,
\label{eq:FSint-dy}
\ee
Notice the similarity between ${\cal F}^S_{\rm DY}$ and ${\cal F}^S_{\rm DIS}$,
Eq.~(\ref{eq:cal-F-S-DIS}).

\subsection{\label{sec:gen:ee-coll} The Collins azimuthal asymmetry in
$e^+e^-\to h_1 h_2\,X$ processes}

We finally consider the Collins azimuthal asymmetry for two almost
back-to-back hadrons produced in opposite jets in $e^+e^-$ annihilations.
We do this in the so-called hadronic-plane method, which is the most reliable
from the experimental point of view, since it does not require the
reconstruction of the jet thrust axis.
On the other hand, from the theoretical point of view this method
explicitly requires the assumption of a factorised, Gaussian-shaped
transverse momentum dependence in the fragmentation functions.
The adoption of the thrust-axis method, which is somehow
more clean theoretically, would lead to similar results and conclusions.

 In the hadronic-plane kinematical configuration,
one measures the transverse momentum, $\bfP_{1T}$, of the first hadron,
$h_1$, w.r.t. the plane containing the initial lepton beams and the second
hadron $h_2$. Following Ref.~\cite{Anselmino:2015sxa} and references therein,
the differential cross section for the process under consideration can be
written as
\be
\dfrac{d\sigma^{e^+e^-\to h_1 h_2\,X}}{dz_1 dz_2 d^2\bm{P}_{1T} d\cos\theta} =
\frac{3\pi\alpha^2}{2s}\,\left\{ D_{h_1h_2} + N_{h_1h_2}\,\cos(2\phi_1)\right\}\,,
\label{eq:dsig-ee}
\ee
where $\theta$ is the angle between the direction of motion of $h_2$ and the
beam axis; $\phi_1$ is the azimuthal angle of $\bfP_{1T}$; $z_{1,2}$ are the
light-cone momentum fractions of the hadrons $h_{1,2}$.

{}From Eqs.~(30) and (31) of Ref.~\cite{Anselmino:2015sxa} we have
\bea
D_{h_1h_2} &=& (1+\cos^2\theta)\,\sum_q e_q^2\,D_{h_1/q}(z_1)\,D_{h_2/\bar q}(z_2)\,
\dfrac{\exp\left[ -P_{1T}^2/\langle \tilde{p}_\perp^2\rangle
\right]}{\pi\,\langle \tilde{p}_\perp^2\rangle}\,,
\label{eq:Dhh}\\
N_{h_1h_2} &=& \frac{1}{4}\,\frac{z_1z_2}{z_1^2+z_2^2}\,\sin^2\theta\,
\sum_q e_q^2\,\Delta^N D_{h_1/q^\uparrow}(z_1)\,
\Delta^N D_{h_2/\bar q^\uparrow}(z_2)
\, \dfrac{2\,e\,P_{1T}^2}{\langle \tilde{p}_\perp^2\rangle+\tilde{M}_C^2}\,
\dfrac{\exp\left[ -P_{1T}^2\left( \frac{1}{\tilde{M}_C^2}+\frac{1}
{\langle \tilde{p}_\perp^2\rangle}
\right) \right]}{\pi\,\langle \tilde{p}_\perp^2\rangle}\,,
\label{eq:Nhh}
\eea
where
\be
\tilde{M}_C^2 = \frac{z_1^2+z_2^2}{z_2^2}\,M_C^2 \qquad\qquad
\langle \tilde{p}_\perp^2\rangle = \frac{z_1^2+z_2^2}{z_2^2}\,\avp
\label{eq:pT-mc-tilde}
\ee
and $M_C$ is the parameter introduced in the Collins function,
Eqs.~(\ref{Coll-frag}) and (\ref{S-C-2}). Notice that the factorised
$z$-dependent part of the Collins function, $\Delta^N D_{h/q^\uparrow}(z)$,
was denoted $\tilde{\Delta}^N D_{h/q^\uparrow}(z)$ in
Ref.~\cite{Anselmino:2015sxa}.

For simplicity, we are assuming that $h_1$, $h_2$ are both either pions or kaons,
leaving aside for instance the $\pi K$ case that would in general require two different
$\langle p_\perp^2\rangle$ values.

The azimuthal asymmetries of interest are the $\cos(2\phi_1)$ modulations
of the cross section (\ref{eq:dsig-ee}), driven by the ratios
$N_{h_1h_2}/D_{h_1h_2}$. Data have been taken for different charge
combinations of the two hadrons, that is: $h_1 h_2 = \pi^+\pi^- + \pi^-\pi^+$ ($U$),
$\pi^+\pi^+ + \pi^-\pi^-$ ($L$) and $\pi^+\pi^- + \pi^-\pi^+ +
\pi^+\pi^+ + \pi^-\pi^-$ ($C$). The actual quantities measured are
\be
A_0^{UL(C)} \simeq P_0^U-P_0^{L(C)}\,,
\label{eq:A-double}
\ee
where
\be
P_0^{U,L,C} = \dfrac{N_{h_1h_2}^{U,L,C}}{D_{h_1h_2}^{U,L,C}}\>\cdot
\label{eq:p0-ulc}
\ee

{}From Eqs.~(\ref{eq:Dhh}) and (\ref{eq:Nhh}) we can write:
\be
P_0^{h_1h_2}(z_1,z_2,P_{1T};\theta) =
A_{\rm ee}^{h_1h_2}(z_1,z_2;\theta)\,F_{\rm ee}^C(z_1,z_2,P_{1T})\,,
\label{eq:p0-af-ee}
\ee
where
\bea
A_{\rm ee}^{h_1h_2}(z_1,z_2;\theta) &=& \dfrac{1}{4}\,
\dfrac{\sin^2\theta}{1+\cos^2\theta}\,
\dfrac{z_1z_2}{z_1^2+z_2^2}\,
\dfrac{\sum_q e_q^2\,\Delta^N D_{h_1/q^\uparrow}(z_1)\,
\Delta^N D_{h_2/\bar q^\uparrow}(z_2)}
{\sum_q e_q^2\,D_{h_1/q}(z_1)\,D_{h_2/\bar q}(z_2)}
\label{eq:Aee-coll}\\
\nonumber\\\nonumber\\
F_{\rm ee}^C(z_1,z_2,P_{1T}) &=&
\dfrac{\dfrac{2\,e\,P_{1T}^2}{\langle \tilde{p}_\perp^2\rangle +
\tilde{M}_C^2}\,\dfrac{\exp\left[ -P_{1T}^2\left( \frac{1}{\tilde{M}_C^2}+
\frac{1}{\langle \tilde{p}_\perp^2\rangle}\right)
\right]}{\pi\,\langle \tilde{p}_\perp^2\rangle}}{\dfrac{\exp\left[ -P_{1T}^2/
\langle \tilde{p}_\perp^2\rangle \right]}{\pi\,\langle \tilde{p}_\perp^2\rangle}}
\,\cdot
\label{eq:Fee-coll}
\eea

Also in this case we proceed by integrating separately the numerator and
denominator of $F_{\rm ee}^C$ over $P_{1T}\, \de P_{1T}$ in the full range
$[0,+\infty)$, finding
\be
P_0^{h_1h_2}(z_1,z_2;\theta) =
A_{\rm ee}^{h_1h_2}(z_1,z_2;\theta)\,{\cal F}_{\rm ee}^C\,,
\label{eq:p0-af-ee-int}
\ee
with
\be
{\cal F}_{\rm ee}^C(\rho_C) = 2\,e\,\rho_C^2(1-\rho_C)\,.
\label{eq:cal-Fee-coll}
\ee
Notice that ${\cal F}_{\rm ee}^C$ is independent of $z_1$, $z_2$.

\subsection{\label{sec:gen:summary} Summary of main formulas}

It is convenient to collect here, all together, the main results of the
previous subsections which will be used in the next Sections.

\subsubsection{ $P_T$-integrated Sivers asymmetry in the SIDIS process
$\ell p^\uparrow\to \ell^\prime h\,X$}
\be
A_{UT}^{\sin(\phi_h-\phi_S)}(x,z) =
A^S_{\rm DIS}(x,z)\,{\cal F}^S_{\rm DIS}(z)
\quad\quad\quad A^S_{\rm DIS}(x,z)\>\, {\rm as \>\, in \>\,
Eq.~(\ref{eq:A-S-DIS})}
\label{eq:A-UT-AF-int-s}
\ee
\be
{\cal F}^S_{\rm DIS}(z,\rho_S,\xi_1) = \sqrt{\frac{e\pi}{2}}\,\left[\,
\dfrac{\rho_S^3(1-\rho_S)}{\rho_S + \xi_{1}/z^2}\right]^{1/2}
\quad\quad\quad
\xi_1 = \dfrac{\avp}{\avk}\quad\quad\quad
\rho_S = \frac{\avk_S}{\avk} = \dfrac{1}{1+\dfrac{\avk}{M_S^2}}
\label{eq:S-sidis}
\ee

\subsubsection{$P_T$-integrated Collins asymmetry in the SIDIS process
$\ell p^\uparrow\to \ell^\prime h\,X$}
\be
A_{UT}^{\sin(\phi_h+\phi_S)}(x,y,z) = A^C_{\rm DIS}(x,y,z)
\,{\cal F}^C_{\rm DIS}(z)
\quad\quad\quad A^C_{\rm DIS}(x,z)\>\, {\rm as \>\, in \>\,
Eq.~(\ref{eq:A-S-DIS-coll})}
\label{eq:A-UT-AF-coll-int-s}
\ee
\be
{\cal F}^C_{\rm DIS}(z,\rho_C,\xi_1/\xi_T) = \sqrt{\frac{e\pi}{2}}\,\left[\,
\dfrac{\rho_C^3(1-\rho_C)}{\rho_C + z^2(\xi_{T}/\xi_1)}\right]^{1/2}
\quad\quad\quad
\xi_T = \frac{\avk_T}{\avk} \quad\quad\quad
\rho_C = \frac{\avp_C}{\avp} = \dfrac{1}{1+\dfrac{\avp}{M_C^2}}
\label{eq:C-sidis}
\ee

\subsubsection{ $q_T$-integrated Sivers asymmetry in the DY process,
$h_1^\uparrow h_2\to \ell^+\ell^-\,X$}
\be
A_N^{\rm DY}(y,M) = A^S_{\rm DY}(x_1,x_2)\,{\cal F}^S_{\rm DY}
\quad\quad\quad A^S_{\rm DY}(x_1,x_2)\>\, {\rm as \>\, in \>\,
Eq.~(\ref{eq:S-dy})}
\label{eq:A-DY-AF-int-s}
\ee
\be
{\cal F}^S_{\rm DY}(\rho_S,\xi_{21}) = \sqrt{\frac{e\pi}{2}}\,\left[\,
\dfrac{\rho_S^3(1-\rho_S)}{\rho_S+\xi_{21}}\right]^{1/2}
\quad\quad\quad
\xi_{21} = \dfrac{\langle k_{\perp 2}^2\rangle}{\langle k_{\perp 1}^2\rangle}
\quad\quad\quad
\rho_S = \frac{\avk_S}{\langle k_{\perp 1}^2\rangle} =
\dfrac{1}{1+\dfrac{\langle k_{\perp 1}^2\rangle}{M_S^2}}
\label{eq:S-DY}
\ee

\subsubsection{$P_T$-integrated Collins asymmetry in the process
$e^+e^-\to h_1h_2\,X$}
\be
P_0^{h_1h_2}(z_1,z_2;\theta) =
A_{\rm ee}^{h_1h_2}(z_1,z_2;\theta)\,{\cal F}_{\rm ee}^C
\quad\quad\quad A_{\rm ee}^{h_1h_2}(z_1,z_2;\theta)\>\, {\rm as \>\, in \>\,
Eq.~(\ref{eq:Aee-coll})}
\label{eq:p0-af-ee-int-s}
\ee
\be
{\cal F}_{\rm ee}^C(\rho_C) = 2\,e\,\rho_C^2(1-\rho_C)\,.
\label{eq:C-ee}
\ee

\section{\label{sec:siv} The Sivers case}
In this section we consider the fit of the Sivers SSAs in SIDIS
and Drell-Yan processes, and the possible phenomenological uncertainties
induced by the strong correlation between $\avk$ and $\avp$ in SIDIS azimuthal
asymmetries, Eqs.~(\ref{PhT}) and (\ref{eq:PT-S}). More precisely, since at
present only a few experimental results are available on the Sivers SSA in
Drell-Yan processes, we study the consequences for predictions on this observable due to the uncertainty on $\avk$ and $\avp$ as extracted from SIDIS data.

According to the present experimental situation, the amount of available SIDIS
data on the Sivers azimuthal asymmetry $A_{UT}^{\sin(\phi_h-\phi_S)}$,
allows to obtain a sufficiently well-constrained parameterisation of the quark
Sivers distributions, $\Delta f_{q/\pup}(x)$, at least in some kinematical
ranges (the present SIDIS data are limited to the $x_{\rm B} \lesssim 0.3$
region). We denote by $\hat{\rho}_S$ and $\hat{\xi}_1$ the particular values
of $\rho_S$ and $\xi_1$, Eq.~(\ref{eq:S-sidis}), corresponding to a SIDIS best
fit of the Sivers function. We shall adopt the ``hat" symbol also for the
corresponding Sivers SSAs.

Notice that for Drell-Yan processes with two different initial beams, as it is
the case for the COMPASS experiment at CERN, where one considers the reaction
$\pi \, p^\uparrow\to \ell^+\ell^-\,X$, one should also take into account the
parameter $\xi_{21}$ introduced in the previous sections. In order to simplify
the analysis and focus on the main issue, we only consider the case $\xi_{21}
= 1$, corresponding to $p\,p$ collisions.

As mentioned in the introduction, different studies of unpolarised azimuthal
distributions~\cite{Anselmino:2005nn}, hadron multiplicities~\cite{Signori:2013mda,%
Anselmino:2013lza,Bacchetta:2017gcc}
and the Sivers SSA in SIDIS processes have been performed. These studies have
indeed shown a strong correlation between $\avk$ and $\avp$, which manifests itself
in large differences in the values of $\xi_1 = \avp/\avk$, which can be associated to
different, equally good fits of the same quantities, in particular the Sivers asymmetry
$A_{UT}^{\sin(\phi_h-\phi_S)}$.

To be definite, we consider in particular two different parameterisation sets for
the Sivers distributions, which lead to comparable values of $\chi^2_{\rm dof}$:
\begin{itemize}
\item
The fit of Ref.~\cite{Anselmino:2008sga}, referred to as FIT09, for which
\begin{equation}
\langle k_\perp^2\rangle = 0.25\, {\rm GeV}^2,\qquad \langle p_\perp^2\rangle
= 0.20\, {\rm GeV}^2,
\qquad M_S^2 = 0.34\, {\rm GeV}^2\,,
\label{eq:fit09-1}
\end{equation}
implying
\begin{equation}
\hat{\xi}_1^{(09)} = 0.80,\qquad \hat{\rho}_S^{(09)} = 0.58\,.
\label{eq:fit09-2}
\end{equation}

The complete list of parameters fixing the Sivers functions can be found in
Table~1 of Ref.~\cite{Anselmino:2008sga}, where
more details on the fitting procedure, the parameter extraction and additional references are given.

It is important to remind here that, for this as well as
for all the following reference fits
adopted, the values of $\langle k_\perp^2\rangle$
and $\langle p_\perp^2 \rangle$ are first extracted from observables depending
only on the unpolarised TMD distribution and fragmentation functions, and then used,
as fixed parameters, in the fitting procedure of the azimuthal spin asymmetries.
\item
The fit from Ref.~\cite{Anselmino:2016uie}, referred to as FIT16, for which
\begin{equation}
\langle k_\perp^2\rangle = 0.57\, {\rm GeV}^2,\qquad \langle p_\perp^2\rangle
= 0.12\, {\rm GeV}^2,
\qquad M_S^2 = 0.80\, {\rm GeV}^2\,,
\label{eq:fit16-1}
\end{equation}
implying
\begin{equation}
\hat{\xi}_1^{(16)} = 0.21,\qquad \hat{\rho}_S^{(16)} = 0.58\,.
\label{eq:fit16-2}
\end{equation}
Again, detailed information and the complete list of parameters can be found in
Ref.~\cite{Anselmino:2016uie} and its Table~1.
\end{itemize}

Notice that the two parameterisations show very different values of $\hat{\xi}_1$,
but almost identical values of $\hat{\rho}_S$. This has the consequence that
$\hat{{\cal F}}^S_{\rm DY} \equiv {\cal F}^S_{\rm DY}(\hat{\rho}_S,\xi_{21}=1)$
is the same for the two reference fits, FIT09 and FIT16.

The possibility of obtaining equally good fits of the SIDIS Sivers data with
different values of the parameters, in particular $\xi_1$, can be formalised
by assuming that, at least in some limited regions of the $(\rho_S,\xi_1)$
parameter space, moving away from the reference point along some trajectory,
one keeps having:
\be
A^S_{\rm DIS}(x,z)\,{\cal F}^S_{\rm DIS}(z,\rho_S,\xi_1) \simeq
\hat{A}^S_{\rm DIS}(x,z)\,\hat{{\cal F}}^S_{\rm DIS}(z,\hat{\rho}_S,\hat{\xi}_1)\,.
\label{eq:asfs-fix}
\ee

Notice that by changing the values of $\xi_1$ and $\rho_S$ one obtains in general different
values of ${\cal F}^S_{\rm DIS}(z,\rho_S,\xi_1)$: then, by fitting the same data
either with the l.h.s. or the r.h.s. of Eq.~(\ref{eq:asfs-fix}), one extracts
different values of $\Delta f_{q/\pup}(x)$, which is contained in $A^S_{\rm DIS}$,
Eq.~(\ref{eq:A-S-DIS}). In fact one has:
\be
A^S_{\rm DIS} \simeq \left( \frac{\hat{{\cal F}}^S_{\rm DIS}}
{{\cal F}^S_{\rm DIS}}\right)\hat{A}^S_{\rm DIS}\,.
\label{eq:ASxz}
\ee

The predictions for the DY Sivers asymmetry, made using the SIDIS Sivers
function $\Delta f_{q/\pup}(x)$, are then affected by its uncertainty; as both
$A^S_{\rm DIS}$ and $A^S_{\rm DY}$ are linear in the Sivers function it is
natural to assume that
\be
\frac{A^S_{\rm DY}}{\hat{A}^S_{\rm DY}} \simeq
\frac{A^S_{\rm DIS}}{\hat{A}^S_{\rm DIS}}\,,
\label{eq:ASdysid}
\ee
which, using Eq.~(\ref{eq:ASxz}), implies
\be
A^S_{\rm DY} \simeq \left( \frac{\hat{{\cal F}}^S_{\rm DIS}}
{{\cal F}^S_{\rm DIS}}\right)\,\hat{A}^S_{\rm DY}\,,
\label{eq:ASdy-xz}
\ee

Notice that from Eqs.~(\ref{eq:S-sidis}) and (\ref{eq:ASxz}) one has
\be
{\cal F}^S_{\rm DIS} = R^{S}_{\rm DIS}\, \hat{{\cal F}}^S_{\rm DIS}
\quad\quad A^S_{\rm DIS} \simeq \frac{1}{R^{S}_{\rm DIS}}\,\hat{A}^S_{\rm DIS}\quad\quad {\rm with} \quad\quad
R^{S}_{\rm DIS} = \left[\,\frac{\rho_S^3(1-\rho_S)}{\rho_S+\xi_1/z^2}
\,\frac{\hat{\rho}_S+\hat{\xi}_1/z^2}{\hat{\rho}_S^3(1-\hat{\rho}_S)}\,
\right]^{1/2}\,,
\label{eq:RFDIS}
\ee
and, analogously, from Eq.~(\ref{eq:S-DY}), with $\xi_{21}=1$, and Eq.~(\ref{eq:ASdy-xz}):
\be
{\cal F}^S_{\rm DY} = R^{S}_{\rm DY}\, \hat{{\cal F}}^S_{\rm DY}
\quad\quad A^S_{\rm DY} \simeq \frac{1}{R^{S}_{\rm DIS}}\,\hat{A}^S_{\rm DY}\quad\quad {\rm with} \quad\quad
R^{S}_{\rm DY} = \left[\,\frac{\rho_S^3(1-\rho_S)}{\rho_S+1}
\,\frac{\hat{\rho}_S+1}{\hat{\rho}_S^3(1-\hat{\rho}_S)}\,\right]^{1/2}\,.
\label{eq:RFDY}
\ee

Then, when moving in the parameter space from $(\hat{\rho}_S,\hat{\xi}_1)$ to
$(\rho_S,\xi_1)$ along a generic trajectory,
the predictions for the Sivers DY asymmetry change as:
\be
A^{\rm DY}_N = A^S_{\rm DY} {\cal F}^S_{\rm DY} \simeq
\left(\frac{R^{S}_{\rm DY}}
{R^{S}_{\rm DIS}}\right)\, \hat{A}^S_{\rm DY}\hat{\cal F}^S_{\rm DY} =
R^N_{\rm DY} \hat{A}^{\rm DY}_N\,,
\label{eq:ADY-change}
\ee
where
\be
R^N_{\rm DY} = \left[\,\frac{\rho_S+\xi_1/z^2}{\hat{\rho}_S+\hat{\xi}_1/z^2}
\,\frac{\hat{\rho}_S+1}{\rho_S+1}\,\right]^{1/2}\,.
\label{eq:RNDY}
\ee

Let us now discuss some possible different scenarios, one corresponding to
the parameters of the sets FIT09 and FIT16
($R^{S}_{\rm DY} = 1$),
and two more exploratory cases ($R^{S}_{\rm DIS} = 1$ and $R^N_{\rm DY} = 1$).

\subsection{\label{sec:siv-sc1} Sivers Effect, scenario 1: FIT09 vs. FIT16}

This is the case which motivated our study. We have two different
parameterisation sets of the Sivers distribution, FIT09 and FIT16 discussed
above, which describe comparably well the Sivers azimuthal asymmetry measured
in SIDIS processes. We have investigated to what extent the corresponding
estimates for the Sivers asymmetry in Drell-Yan processes can differ due to
the uncertainty on the $\xi_1$ parameter, Eqs.~(\ref{eq:fit09-2}) and
(\ref{eq:fit16-2}). Notice that, in this case, $\hat{\rho}^{(09)}_S =
\hat{\rho}^{(16)}_S \equiv \hat{\rho}_S$.

{}From Eq.~(\ref{eq:RFDY}) then one sees that $R^{S}_{\rm DY} = 1$
(remember that we are considering the case of $p\,p$ collisions here, that
is $\xi_{21}=1$), and from Eqs.~(\ref{eq:ADY-change}) and (\ref{eq:RNDY})
one obtains that, going from one set of parameters to the other, the
predictions for $A_N^{\rm DY}$ are rescaled as:
\be
A_N^{\rm DY}(\hat{\rho}_S,\hat{\xi}_1^{(16)}) \simeq
\left[\,\frac{\hat{\rho}_S + \hat{\xi}_1^{(16)}/z^2}
{\hat{\rho}_S + \hat{\xi}_1^{(09)}/z^2}
\,\right]^{1/2}\hat{A}^{\rm DY}_N(\hat{\rho}_S,\hat{\xi}_1^{(09)})\,.
\label{eq:ANDY-sc1}
\ee

Using the values given in Eqs.~(\ref{eq:fit09-2}) and (\ref{eq:fit16-2})
one sees that the rescaling factor in the above equation varies from about
0.52 to 0.68 for $z$ in the range $[0.1,0.7]$. Since small $z$ values dominate
the SIDIS data, we find that:
\be
A_N^{\rm DY}(\hat{\rho}_S,\hat{\xi}_1^{(16)}) \simeq \frac 12
\hat{A}^{\rm DY}_N(\hat{\rho}_S,\hat{\xi}_1^{(09)})\,.
\label{eq:ANDY-sc1-num}
\ee

This simple example, based on two available fits of the quark Sivers function,
clearly shows how the uncertainty in the parameter $\xi_1 = \langle p^2_\perp\rangle / \langle k^2_\perp\rangle$, due to the unavoidable strong
correlation between $\langle k^2_\perp\rangle$ and $\langle p^2_\perp\rangle$
in SIDIS processes, Eq.~(\ref{eq:PT-S}), induces large differences when trying
to estimate the Sivers SSA in Drell-Yan processes. This effect should be
carefully taken into account when studying these asymmetries and their related
fundamental properties, like e.g.~the TMD scale evolution of the Sivers function
and its process dependence.

\subsection{\label{sec:siv-sc2} Sivers Effect, scenario 2: fixing
$A^S_{\rm DIS}$ and ${\cal F}^S_{\rm DIS}$}

In the previous scenario, based on the fact that two equally good fits of the
Sivers SIDIS asymmetry, FIT09 and FIT16, yield the same values of $\rho_S$ even if starting with
very different values of $\xi_1$, we have shown how the corresponding
predictions for the Sivers asymmetries in $p\,p$ Drell-Yan processes, can vary
by a factor up to 2, depending on which sets of parameters one uses.
Mathematically, we have kept the validity of Eq.~(\ref{eq:asfs-fix}) by
letting both ${\cal F}^S_{\rm DIS}$ and $A^S_{\rm DIS}$ change, but in opposite
ways (if one decreases, the other increases, and viceversa).

We now extend our investigation of what happens to the estimates for the full
DY Sivers asymmetry if we let the SIDIS parameters vary in different ways in the
$(\rho_S, \xi_1)$ space. We first wonder whether it is possible to keep the
validity of Eq.~(\ref{eq:asfs-fix}) by requiring that both ${\cal F}^S_{\rm DIS}$
and $A^S_{\rm DIS}$ do not change when moving along some lines in the parameter
space (notice that if $A^S_{\rm DIS}$ does not change, then, by Eq.~(\ref{eq:ASdysid}), also $A^S_{\rm DY}$ does not change). {}From Eq.~(\ref{eq:RFDIS}), we see that this request amounts to impose:
\be
R^{S}_{\rm DIS} = \left[\,\frac{\rho_S^3(1-\rho_S)}{\rho_S+\xi_1/z^2}
\,\frac{\hat{\rho}_S+\hat{\xi}_1/z^2}{\hat{\rho}_S^3(1-\hat{\rho}_S)}\,
\right]^{1/2} = 1 \>.
\ee

\begin{figure}[]
\includegraphics[width=8.truecm,angle=0]{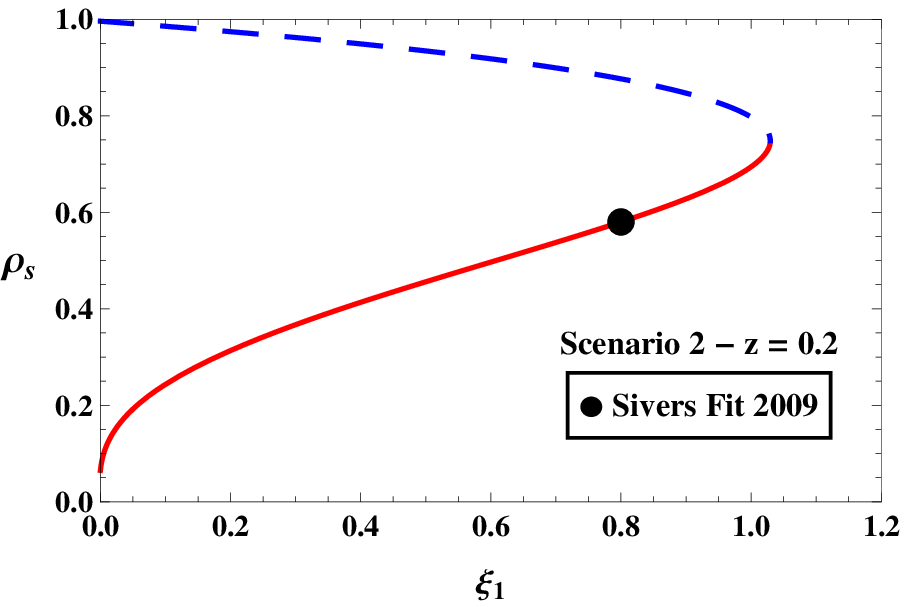}
\includegraphics[width=8.truecm,angle=0]{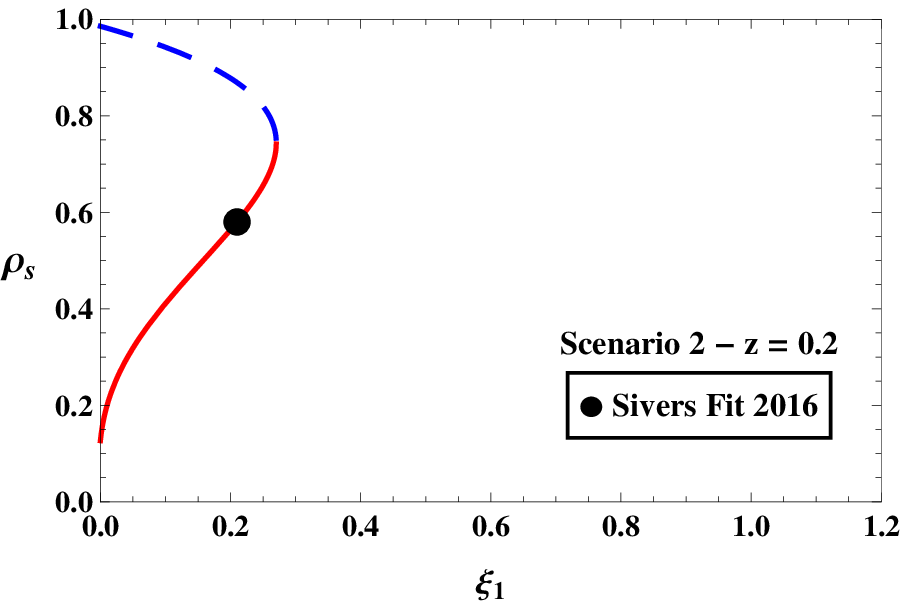}
\caption{The curves in the $(\rho_S,\xi_1$) parameter space show the set
of values of $\rho_S$ and $\xi_1$ which leave unchanged the
$P_T$-integrated factor of the Sivers asymmetry, ${\cal F}^S_{\rm DIS}(z = 0.2)$.
The black dots correspond to the values $\hat{\rho}_S$ and $\hat{\xi}_1$
obtained in the fits of Ref.~\cite{Anselmino:2008sga} (left plot, FIT09) and of
Ref.~\cite{Anselmino:2016uie} (right plot, FIT16), which describe equally well
the SIDIS Sivers asymmetry. Notice that for each value of $\xi_1$ one finds two possible values of $\rho_S$. Similar results are obtained by changing $z$ from
0.2 to 0.4 or 0.6.}
\label{fig:rhoS-vs-xi1-z02}
\end{figure}

\begin{figure}[]
\includegraphics[width=8.truecm,angle=0]{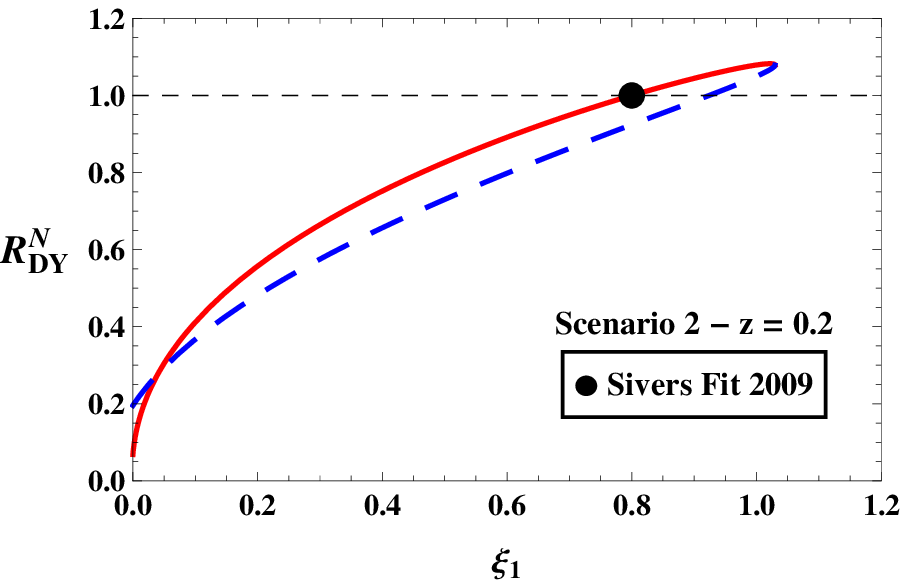}
\includegraphics[width=8.truecm,angle=0]{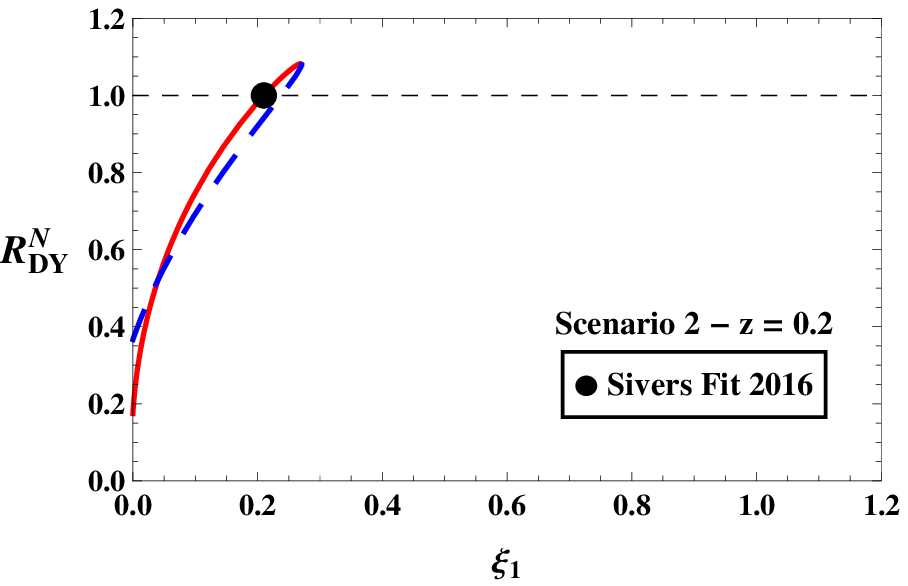}
\caption{The plots show how the predictions for the Drell-Yan
$q_T$-integrated Sivers asymmetry $A_N^{\rm DY}$, Eq.~(\ref{eq:A-DY-AF-int-s}),
change, as functions of $\xi_1$, when the parameters extracted from SIDIS
data move along the corresponding lines of Fig.~\ref{fig:rhoS-vs-xi1-z02}.
The rescaling factor $R^N_{\rm DY}$ is defined in Eqs.~(\ref{eq:ADY-change})
and (\ref{eq:RNDY}). In this scenario the $P_T$-integrated SIDIS Sivers
asymmetry $A_{UT}^{\sin(\phi_h-\phi_S)}$, Eq.~(\ref{eq:A-UT-AF-int-s}), does
not change, together with its factors $A^S_{\rm DIS}$ and ${\cal F}^S_{\rm DIS}$.
}
\label{fig:ratio-FSDY-vs-x1-z02}
\end{figure}

At fixed $\xi_1$ and $z$, this constraint corresponds to a 4th order algebraic
equation in the variable $\rho_S$,
\begin{equation}
\rho_S^4-\rho_S^3+\hat{a}(z)\rho_S + \hat{a}(z)\frac{z^2}{\xi_1} = 0
\quad\quad {\rm with} \quad\quad
\hat{a}(z) = \frac{\hat{\rho}^3_S(1-\hat{\rho}_S)}{\hat{\rho}_S+\hat{\xi}_1/z^2}
\> \cdot
\label{eq:4th-S}
\end{equation}
 and we can look for its (real) solutions in
terms of $\xi_1$ and $z$ in the physical range $0 < \rho_S <1$.
There are in fact two real solutions, at least for some ranges of $\xi_1$ values.
As an example, they are shown, as a function of $\xi_1$ and at fixed $z = 0.2$,
in Fig.~\ref{fig:rhoS-vs-xi1-z02}, respectively for the FIT09 (left panel) and the
FIT16 (right panel) case. The black dots correspond to the position in the
parameter plane of the corresponding reference fit. They both belong to the
lower of the two possible branches of solutions (the red solid and blue
long-dashed curves). The corresponding values of $R^N_{\rm DY}$
($=R^S_{\rm DY}$, in this scenario), that is the
rescaling factor for the predictions of the DY Sivers asymmetry,
Eq.~(\ref{eq:RNDY}), are shown, as a function of $\xi_1$, in
Fig.~\ref{fig:ratio-FSDY-vs-x1-z02}.

The left panel of Fig.~\ref{fig:ratio-FSDY-vs-x1-z02} shows that the rescaling
factor for the set FIT09, for which ${\hat\xi}_1^{(09)} = 0.80$, decreases to
almost 1/2 when $\xi_1$ approaches 0.20, as seen in the previous scenario (notice,
however, that in this case also $\rho_S$ changes). Concerning the set FIT16
(right panel of Fig.~\ref{fig:ratio-FSDY-vs-x1-z02}), we see that, although the
range of $\xi_1$ values leading to an allowed value of $\rho_S$ is more restricted,
in any case the depletion effect on the total DY asymmetry can still be large as
soon as $\xi_1$ decreases.

Notice that, even if our calculation leads to two possible solutions for $\rho_S$
at fixed $\xi_1$ (the reference fits corresponding to the lower one) the rescaling
factor $R^N_{\rm DY}$, which is the quantity of interest from the physical point
of view, is very similar for the two cases. Qualitatively
similar results and conclusions apply when considering $z = 0.4$ and 0.6.

The plots in Figs.~\ref{fig:rhoS-vs-xi1-z02} and \ref{fig:ratio-FSDY-vs-x1-z02}
are shown for all values of $\xi_1$ mathematically compatible with the physical
request $0 < \rho_S < 1$, but one should not forget that very small values of
$\xi_1$ are not realistic. Actually, the range $0.15 \lesssim \xi_1 \lesssim 2.5$
would largely cover most of the parameterisations proposed in the
literature (see also Ref.~\cite{Bacchetta:2017gcc}).

Let us finally stress once more that, as compared to the previous scenario,
in this case it is the $q_T$-integrated component of the overall DY asymmetry
that is rescaled by a factor $R^N_{\rm DY} = R^S_{\rm DY}$,
while the collinear component is approximately unchanged, since $R^S_{\rm DIS} =1$,
see Eqs.~(\ref{eq:RFDY}),~(\ref{eq:ADY-change}).

\subsection{\label{sec:siv-sc3} Sivers Effect, scenario 3: fixing $A^{\rm DY}_N$}

Finally, we wonder whether it is possible to change the parameters $\rho_S$ and
$\xi_1$, moving away from the reference fit values in the parameter space, still
getting the same results not only for the $P_T$-integrated Sivers SIDIS asymmetry,
$A_{UT}^{\sin(\phi_h-\phi_S)}$, but also for the $q_T$-integrated Sivers DY asymmetry, $A^{\rm DY}_N$. This amounts to request:
\be
R^N_{\rm DY} = \left[\,\frac{\rho_S+\xi_1/z^2}{\hat{\rho}_S+\hat{\xi}_1/z^2}
\,\frac{\hat{\rho}_S+1}{\rho_S+1}\,\right]^{1/2} = 1\,,
\label{eq:RN1}
\ee
or, equivalently,
\be
R^{S}_{\rm DY} = R^{S}_{\rm DIS} \>.
\label{eq:dyeqdis}
\ee

By defining
\be
\hat{b}(z) = \frac{\hat{\rho}_S+1}{\hat{\rho}_S+\hat{\xi}_1/z^2}\,,
\label{eq:azdef}
\ee
Eq.~(\ref{eq:RN1}) translates into the simple linear relation
\be
\rho_S = \frac{\hat{b}(z)\,\xi_1/z^2-1}{1-\hat{b}(z)}
\qquad\qquad {\rm for} \qquad \hat{b}(z) \neq 1\>,
\label{eq:rxlin}
\ee
where $\hat{b}(z)$ is a rapidly increasing function of $z$.
To have an idea, $\hat{b}^{(09)}(z=0.2) \simeq 0.08$,
$\hat{b}^{(09)}(z=0.6) \simeq 0.56$,  $\hat{b}^{(16)}(z=0.2) \simeq 0.27$,
 $\hat{b}^{(16)}(z=0.6) \simeq 1.36$.
Requiring that $0 < \rho_S < 1$ restricts the allowed values of $\xi_1$ in
terms of $\hat{b}(z)\,$:
\bea
\mbox{if}& \hat{b}\, < 1,\quad \mbox{then}\quad &
 z^2/\hat{b}\, <\, \xi_1\, <\, z^2\,(2-\hat{b})/\hat{b} \nonumber \\
 &&\label{eq:xirange}\\
\mbox{if}& \hat{b}\, > 1,  \quad \mbox{then}\quad &
 z^2\,(2-\hat{b})/\hat{b}\, <\, \xi_1\, <\, z^2/\hat{b}\,.\nonumber
\eea

\begin{figure}[]
\includegraphics[width=8.truecm,angle=0]{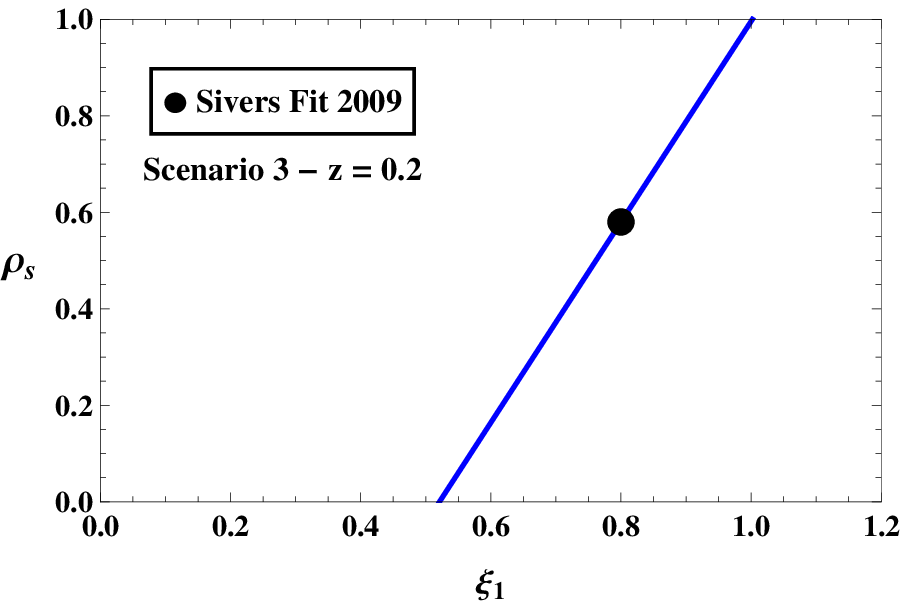}
\includegraphics[width=8.truecm,angle=0]{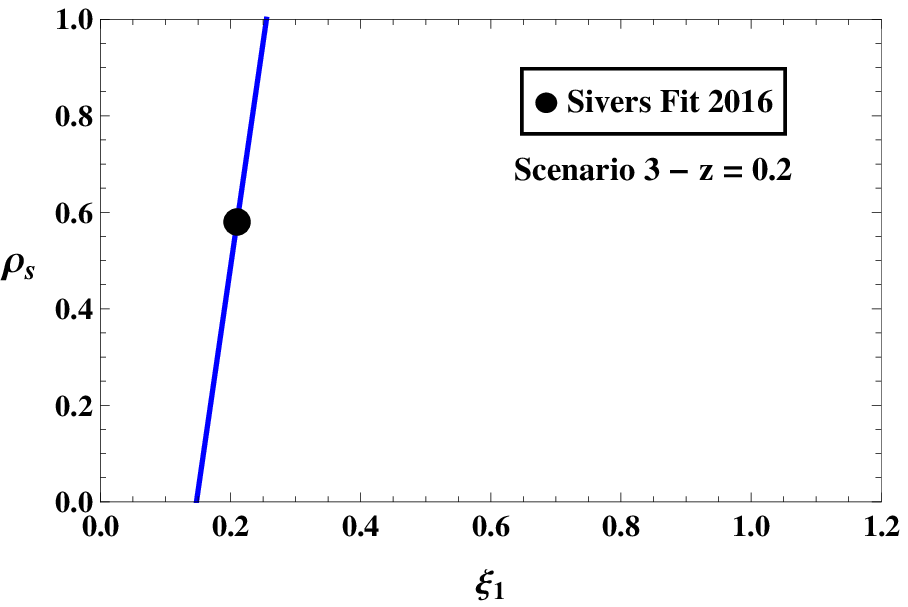}
\caption{The two lines in the $(\rho_S,\xi_1$) parameter space show the set
of values of $\rho_S$ and $\xi_1$ which leave unchanged the SIDIS Sivers
asymmetry $A_{UT}^{\sin(\phi_h-\phi_S)}$, Eq.~(\ref{eq:A-UT-AF-int-s}),
{\it and} the predictions for the Drell-Yan Sivers asymmetry $A^{\rm DY}_N$,
Eq.~(\ref{eq:ADY-change}). The black dots correspond to the values $\hat{\rho}_S$
and $\hat{\xi}_1$ obtained in the fits of Ref.~\cite{Anselmino:2008sga} (left
plot, FIT09) and of Ref.~\cite{Anselmino:2016uie} (right plot, FIT16), which
describe equally well the SIDIS Sivers asymmetry.}
\label{fig:ANDY-fixed}
\end{figure}

\begin{figure}[]
\includegraphics[width=8.truecm,angle=0]{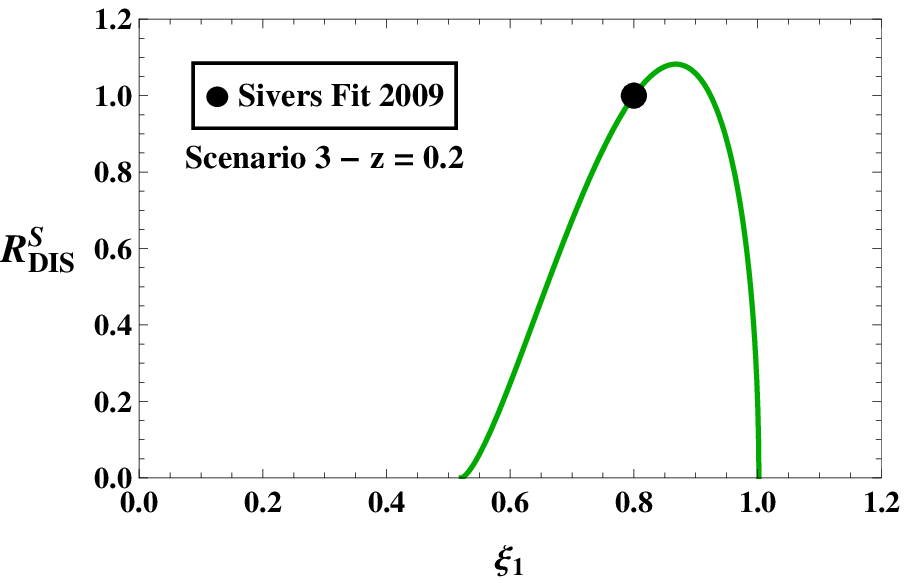}
\includegraphics[width=8.truecm,angle=0]{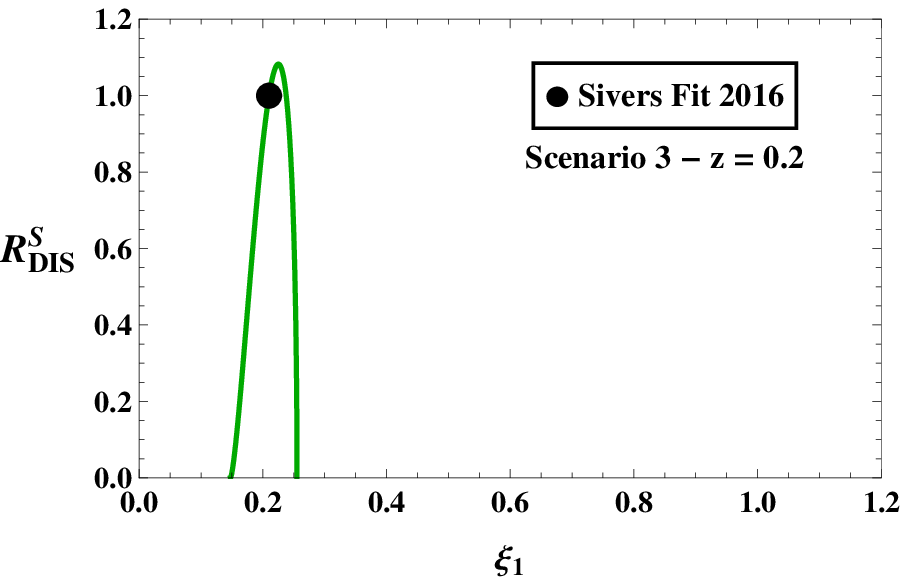}
\caption{The plots show the rescaling factor $R^{S}_{\rm DIS} =
R^{S}_{\rm DY}$ which fixes the changes of ${\cal F}^S_{\rm DIS}$ and
${\cal F}^S_{\rm DY}$, Eqs.~(\ref{eq:RFDIS}) and (\ref{eq:RFDY}), when the
parameters $\rho_S$ and $\xi_1$ move as in the corresponding plots of Fig.~\ref{fig:ANDY-fixed}.
Notice that $A^S_{\rm DIS}$ and $A^S_{\rm DY}$,
and therefore the collinear component of the Sivers function,
change as $1/R^{S}_{\rm DIS}$. In this scenario both the $P_T$-integrated
SIDIS Sivers asymmetry $A_{UT}^{\sin(\phi_h-\phi_S)}$,
Eq.~(\ref{eq:A-UT-AF-int-s}), and the $q_T$-integrated DY Sivers asymmetry
$A_N^{\rm DY}$, Eq.~(\ref{eq:A-DY-AF-int-s}), do not change.}
\label{fig:R-fixed}
\end{figure}

As an example, Fig.~\ref{fig:ANDY-fixed} shows, for the sets FIT09 (left panel)
and FIT16 (right panel) and for $z=0.2$, the values of $\rho_S$ corresponding
to the allowed $\xi_1$ range, that keep fixed the values of both the total
Drell-Yan and SIDIS Sivers asymmetries when moving away from the $(\hat{\rho}_S,
\hat{\xi}_1)$ values of the corresponding fit.

The slope of the straight lines in the plots increases, and therefore the
allowed range for $\xi_1$ shrinks, as $\hat{b}(z)$
approaches 1, changing sign when it crosses this value. For $\hat{b}(z) = 1$,
that is for $0 < z= \hat{\xi}_1^{\;1/2} < 1$, Eq.~(\ref{eq:RN1}) can be fulfilled
only for $\xi_1 \equiv \hat{\xi}_1$ and $\rho_S$ is undetermined.

Notice that in this case, like in scenario 1, although the total Sivers
asymmetries are unchanged, the separate factors depending respectively
on the longitudinal momentum fractions and on the transverse momenta,
$A^S$ and ${\cal F}^S$, change according to Eqs.~(\ref{eq:RFDIS}),
(\ref{eq:RFDY}) and (\ref{eq:dyeqdis}). The rescaling factor
$R^{S}_{\rm DIS} = R^{S}_{\rm DY}$ is shown in Fig.~\ref{fig:R-fixed}
as a function of $\xi_1$ in the allowed range, for the two reference fits.

By comparing Fig.~\ref{fig:rhoS-vs-xi1-z02} and Fig.~\ref{fig:ANDY-fixed} we
see that future $q_T$-integrated data on the Drell-Yan Sivers asymmetries could
constrain more severely the range of allowed $\xi_1$ values, in particular for
the FIT16 set (right panel of Fig.~\ref{fig:ANDY-fixed}). On the other hand,
Fig.~\ref{fig:R-fixed} shows that even in this restricted range, as
soon as the value of $\xi_1$ changes (with respect to that of the reference fits)
the two factors of the total asymmetry, $A^S$ which depend on the longitudinal
momentum fractions and ${\cal F}^S$ which depends on the transverse momenta,
can change by a sizeable factor. For $A^S$ this implies a sizeable change
in the collinear part of the Sivers distribution function $\Delta^N f_{q/\pup}(x)$, like
$1/R^S_{\rm DIS}$.

Again, the plots shown in Fig.~\ref{fig:R-fixed} cover all variable ranges
mathematically allowed, but one should keep in mind that too small values
of $R^{S}_{\rm DIS}$ are not physically acceptable. Such values would
yield large values of $A^{S}_{\rm DIS}$ and $A^{S}_{\rm DY}$ (see
Eqs.~(\ref{eq:RFDIS}) and (\ref{eq:RFDY})) and, consequently, large values of
$\Delta f_{q/\pup}(x)$ which eventually violate the positivity bound $|\Delta f_{q/\pup}(x)|
\leq 2\,f_{q/p}(x)$.

These results, and those of the previous two scenarios, clearly show how the
choice of a specific set for the Gaussian widths of the unpolarised TMDs could play a crucial role
in the extraction of the Sivers function from the analysis of the corresponding SIDIS azimuthal asymmetries
and, as a consequence, in the predictions for the Sivers asymmetries in DY processes.

\section{\label{sec:col} The Collins case}

Let us now extend the considerations of the previous Section
to the Collins asymmetries, and see how the
uncertainty on the choice of $\xi_1$ can affect the extraction of the
transversity distribution and the Collins function from SIDIS and $e^+e^-$
annihilation data, Eqs.~(\ref{eq:A-UT-AF-coll-int-s}), (\ref{eq:C-sidis})
and~(\ref{eq:p0-af-ee-int-s}), (\ref{eq:C-ee}). Notice that, although
${\cal F}_{\rm ee}^C(\rho_C)$ does not depend explicitly on $\xi_1$, possible
conditions on ${\cal F}_{\rm DIS}^C$ induce a correlation between $\rho_C$ and
$\xi_1$.

The Collins case is more complicated than the Sivers case.
In fact, in the latter case the explored
Sivers function always enters linearly,
convoluted either with the unpolarised FF function in the SIDIS asymmetries,
or with the unpolarised PDF in the DY asymmetries.
Instead, in the Collins asymmetries, the Collins FF enters linearly in the SIDIS case --
coupled to the transversity distribution -- while in the $e^+e^-$ case it appears ``quadratically'',
in the sense that the Collins function associated to hadron $h_1$ is convoluted
with the Collins function associated to hadron $h_2$. This makes the analysis less direct, since
variations in the transverse momentum dependent factors can
generate different effects
on the $x$ and $z$ dependent parts.
More precisely, they can affect only the
transversity distribution $h_1^q(x)$, or only the collinear part of the Collins
FF $\Delta^N D_{h/q^\uparrow}(z)$, or both of them simultaneously.

Moreover, no experimental data are presently available on the $p_\perp$ distributions
in the cross section for $e^+e^- \to h_1 h_2\, X$ processes, from which one could
attempt an extraction of the unpolarized $\avp$ width. Old measurements
exist, that were recently analysed in Ref.~\cite{Boglione:2017jlh}, but
they correspond to single hadron production in $e^+e^-$ annihilations, a
process for which TMD factorisation theorems are not proven to be
applicable.

In order to simplify our discussion, we assume that changes in the values
of $\xi_1$ and $\rho_C$ will possibly reflect only in variations of the overall
numerical factors appearing in the collinear parts of the transversity
distribution and the Collins FF, rather than in their functional shapes.
At the qualitative level of the present treatment, this allows to focus on
the main effects avoiding additional complications. For the same reason,
we take $\xi_T = \langle k_\perp^2\rangle_T/\langle k_\perp^2\rangle = 1$,
assuming that the transversity distribution has the same transverse momentum
dependence as the unpolarised TMDs.

As for the Sivers case, in our analysis we consider two different reference
parameterisations for the transversity distribution and the Collins FF with
comparable accuracies of the corresponding fits to data:
\begin{itemize}
\item
The fit of Ref.~\cite{Anselmino:2007fs}, referred to as FIT07 in the following,
for which
\be
\langle k_\perp^2\rangle = 0.25\, {\rm GeV}^2,\qquad \langle p_\perp^2\rangle
= 0.20\, {\rm GeV}^2,
\qquad M_C^2 = 0.88\, {\rm GeV}^2\,,
\label{eq:fit07-1}
\ee
implying
\be
\hat{\xi}_1^{(07)} = 0.80,\qquad \hat{\rho}_C^{(07)} = 0.81\,.
\label{eq:fit07-2}
\ee
The complete list of parameters can be found in Table~II of
Ref.~\cite{Anselmino:2007fs}, where
more details on the fitting procedure, the parameter extraction and additional references are given.

\item
The fit from Ref.~\cite{Anselmino:2015sxa}, referred to as FIT15, for which
\be
\langle k_\perp^2\rangle = 0.57\, {\rm GeV}^2,\qquad \langle p_\perp^2\rangle
= 0.12\, {\rm GeV}^2,
\qquad M_C^2 = 0.28\, {\rm GeV}^2\,,
\label{eq:fit15-1}
\ee
corresponding to
\be
\hat{\xi}_1^{(15)} = 0.21,\qquad \hat{\rho}_C^{(15)} = 0.70\,.
\label{eq:fit15-2}
\ee
Again, full details and the complete list of parameters can be found in Ref.~\cite{Anselmino:2015sxa} and its Table~I.
\end{itemize}

We recall that for the Collins asymmetry in $e^+e^-$ annihilations we have
considered here the $A_0$ asymmetry, corresponding to the experimental
``hadronic-plane" setup, where no direct reference to the $q\bar q$ jet
thrust axis is made (see Ref.~\cite{Anselmino:2015sxa} and references therein).
Notice also that, similarly to the SIDIS Sivers case, the two reference fits
differ significantly in the values of $\hat{\xi}_1$ and much less in the
values of $\hat{\rho}_C$ ($\hat{\rho}_S$ for the Sivers asymmetry).

In the Sivers case discussed in the previous Section, we investigated how
the freedom left on the parameters $\xi_1$ and $\rho_S$ by SIDIS data could
affect the predictions for the Sivers asymmetry in DY processes. This was
because of the lack of experimental information
on polarised DY scattering experiments.
In the case of the
Collins asymmetry, instead, sufficient experimental information is available
both from SIDIS and $e^+e^-$ annihilation data. We then investigate the
freedom left on the parameters $\xi_1$ and $\rho_C$ by these data; that is,
we study whether, moving in the parameter space $(\rho_C,\xi_1)$ away from
a given reference set $(\hat{\rho}_C,\hat{\xi}_1)$, the following relations remain true:
\bea
A^C_{\rm DIS}{\cal F}^C_{\rm DIS}(\rho_C,\xi_1) &\simeq&
\hat{A}^C_{\rm DIS}\hat{{\cal F}}^C_{\rm DIS}(\hat{\rho}_C,\hat{\xi}_1)\,,
\nonumber \\
A^C_{\rm ee}{\cal F}^C_{\rm ee}(\rho_C) &\simeq&
\hat{A}^C_{\rm ee}\hat{{\cal F}}^C_{\rm ee}(\hat{\rho}_C)\,.
\label{eq:AC-fix}
\eea

Notice that by using Eqs.~(\ref{eq:C-sidis}), with $\xi_T =1$, and
(\ref{eq:C-ee}), in complete analogy with
the Sivers case, we can also write:
\be
{\cal F}^C_{\rm DIS} = R^{C}_{\rm DIS}\, \hat{{\cal F}}^C_{\rm DIS}
\quad\quad A^C_{\rm DIS} \simeq \frac{1}{R^{C}_{\rm DIS}}\,\hat{A}^C_{\rm DIS}
\quad\quad {\rm with} \quad\quad
R^C_{\rm DIS} = \left[\,\frac{\rho_C^3(1-\rho_C)}{\rho_C+z^2/\xi_1}\,
\frac{\hat{\rho}_C+z^2/\hat{\xi}_1}{\hat{\rho}_C^3(1-\hat{\rho}_C)}\,\right]^{1/2}\,,
\label{eq:RCDIS}
\ee
and
\be
{\cal F}^C_{\rm ee} = R^{C}_{\rm ee}\, \hat{{\cal F}}^C_{\rm ee}
\quad\quad A^C_{\rm ee} \simeq \frac{1}{R^{C}_{\rm ee}}\,\hat{A}^C_{\rm ee}
\quad\quad {\rm with} \quad\quad
R^C_{\rm ee} = \frac{\rho_C^2(1-\rho_C)}{\hat{\rho}_C^2(1-\hat{\rho}_C)}\>\cdot
\label{eq:RCee}
\ee

Due to the factorised nature of our approach, there could be several solutions
of Eqs.~(\ref{eq:AC-fix}). We consider, as examples, a few possible scenarios
which differ by one further additional condition,
leading to different ways of modifying the collinear and transverse-momentum dependent terms in the
asymmetries and, ultimately, the corresponding components of the transversity distribution
$h_1^q(x)$ and of the Collins FF $\Delta^N D_{h/q^\uparrow}(z)$.

\subsection{\label{sec:col-sc1} Collins Effect, scenario 1}

In this scenario we look for possible allowed sets of ($\rho_C, \xi_1$) values
which not only leave unchanged the two (SIDIS and $e^+e^-$) Collins asymmetries,
Eq.~(\ref{eq:AC-fix}), but also the $P_T$-integrated SIDIS Collins factor
${\cal F}^C_{\rm DIS}$:
\be
{\cal F}^C_{\rm DIS}(\rho_C,\xi_1) = \hat{{\cal F}}^C_{\rm DIS}(\hat{\rho}_C,\hat{\xi}_1)\>,
\label{eq:FC-fix1}
\ee
that is:
\be
R^C_{\rm DIS} = \left[\,\frac{\rho_C^3(1-\rho_C)}{\rho_C+z^2/\xi_1}\,
\frac{\hat{\rho}_C+z^2/\hat{\xi}_1}{\hat{\rho}_C^3(1-\hat{\rho}_C)}\,
\right]^{1/2} = 1 \>.
\label{eq:RC-sc1}
\ee

\begin{figure}[]
\includegraphics[width=8.truecm,angle=0]{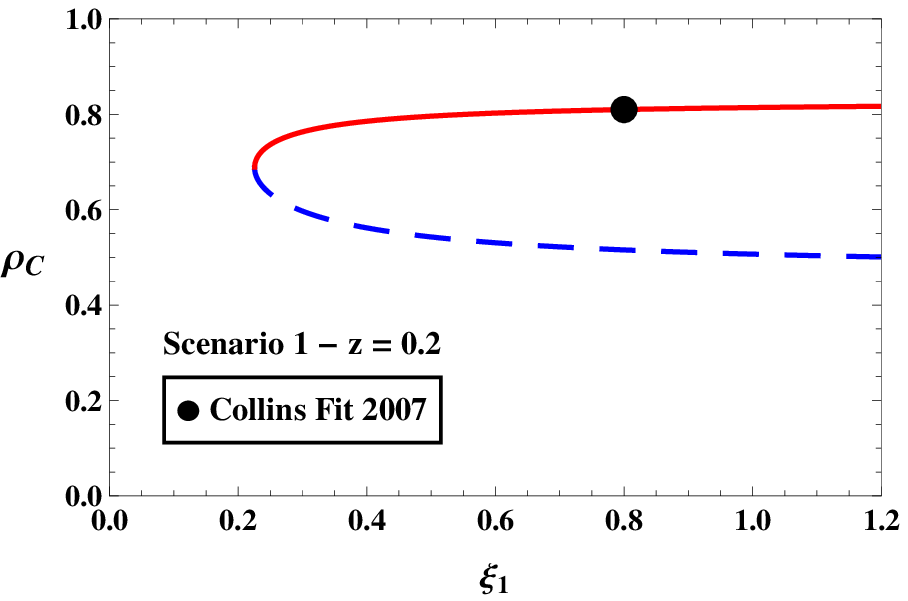}
\includegraphics[width=8.truecm,angle=0]{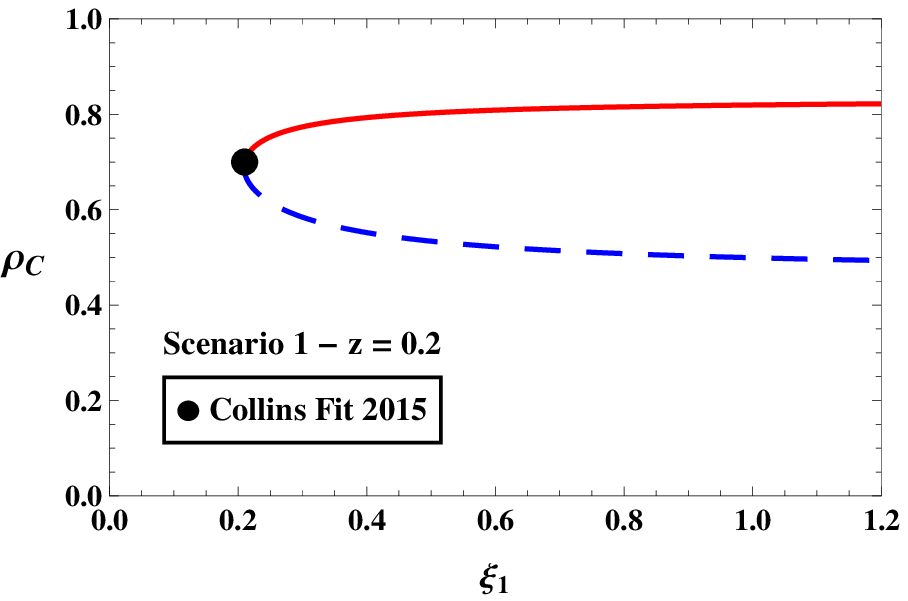}
\caption{The curves in the $(\rho_C,\xi_1$) parameter space show the set
of values of $\rho_C$ and $\xi_1$ which leave unchanged the $P_T$-integrated
factor of the Collins asymmetry, ${\cal F}^C_{\rm DIS}(z = 0.2)$. The black dots
correspond to the values $\hat{\rho}_C$ and $\hat{\xi}_1$ obtained in the
fits of Ref.~\cite{Anselmino:2007fs} (left plot, FIT07) and of
Ref.~\cite{Anselmino:2015sxa} (right plot, FIT15), which describe equally well
the SIDIS and $e^+e^-$ Collins asymmetries. Notice that for each value of $\xi_1$
one finds two possible values of $\rho_C$. Similar results are obtained
by changing $z$ from 0.2 to 0.4 or 0.6.}
\label{fig:C-vs-xi1-z02-sc1}
\end{figure}

As for the Sivers case, the above constraint corresponds to a 4th order
algebraic equation for $\rho_C$, at fixed $\hat{\xi}_1$, $\hat{\rho}_C$ and $z$:
\be
\rho_C^4-\rho_C^3+\hat{c}(z)\rho_C + \hat{c}(z)\frac{z^2}{\xi_1} = 0
\quad\quad {\rm with} \quad\quad
\hat{c}(z) = \frac{\hat{\rho}^3_C(1-\hat{\rho}_C)}{\hat{\rho}_C+z^2/\hat{\xi}_1}
\> \cdot \label{eq:4th-C}
\ee
Again, it turns out that 2 of the 4 possible solutions for $\rho_C$ are complex,
while the other two can be real, at least for some range of $\xi_1$ values.
As an example, they are shown in Fig.~\ref{fig:C-vs-xi1-z02-sc1},
for both the FIT07 (left panel) and FIT15 (right panel) parameterisations,
as a function of $\xi_1$ at fixed $z = 0.2$. Notice that, although the plots
are shown up to $\xi_1=1.2$, at variance with $\rho_S$ for the Sivers case,
the two solutions for $\rho_C$ survive, almost constant, up to much larger
$\xi_1$ values.

{}From Eqs.~(\ref{eq:RCDIS}) and (\ref{eq:RCee}) we have, in this scenario,
\be
A^C_{\rm DIS} = \hat{A}^C_{\rm DIS} \quad\quad
A^C_{\rm ee} =  \frac{1}{R^{C}_{\rm ee}}\,\hat{A}^C_{\rm ee} \>.
\label{AC-sc-1}
\ee

\begin{figure}[]
\includegraphics[width=8.truecm,angle=0]{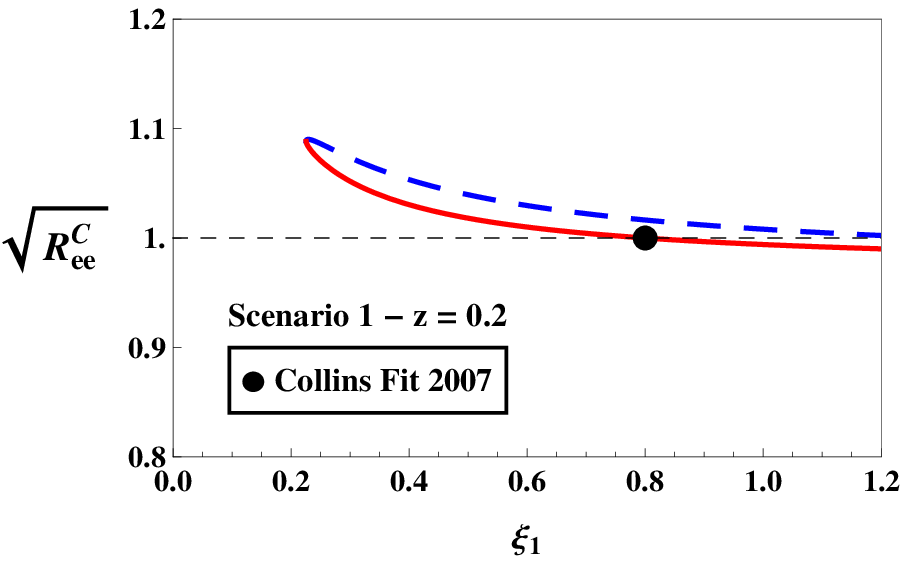}
\includegraphics[width=8.truecm,angle=0]{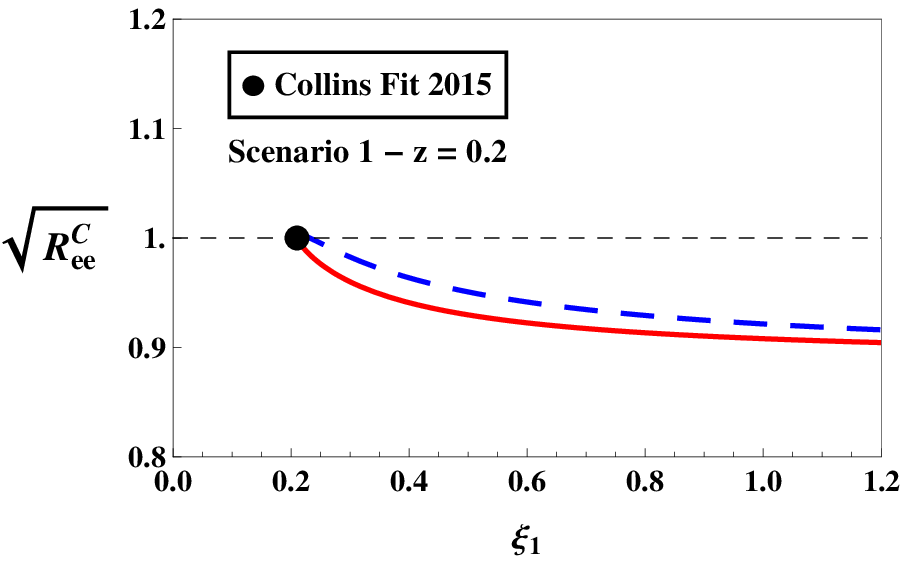}
\caption{The expected rescaling factor $\sqrt{R^C_{\rm ee}}$ for the collinear
transversity distribution $h_1^q(x)$ when the parameters $\rho_C,\,\xi_1$ move
away from the reference fit values as in the corresponding plots of
Fig.~\ref{fig:C-vs-xi1-z02-sc1}. Simultaneously, the Collins collinear
distribution $\Delta^N D_{h/q^\uparrow}(z)$ rescales as $1/\sqrt{R^C_{\rm ee}}$.
In this scenario the total $P_T$-integrated Collins asymmetries,
Eqs.~(\ref{eq:A-DY-AF-int-s}) and (\ref{eq:p0-af-ee-int-s}), remain unchanged,
as well as the ${\cal F}^C_{\rm DIS}$ factor.}
\label{fig:C-REcc-vs-xi-z02-sc1}
\end{figure}

Let us remind that $A^C_{\rm DIS}$, Eq.~(\ref{eq:A-S-DIS-coll}), is a linear
convolution of the transversity distribution $h_1^q(x)$ and the collinear component of the Collins
function $\Delta^N D_{h/q^\uparrow}(z)$, while $A^C_{\rm ee}$,
Eq.~(\ref{eq:Aee-coll}), is ``quadratic" in $\Delta^N D_{h/q^\uparrow}$. Then,
it is reasonable to assume that, in order to keep satisfying Eqs.~(\ref{AC-sc-1})
while the parameters $(\rho_c,\xi_1)$ vary as in Fig.~\ref{fig:C-vs-xi1-z02-sc1},
$\Delta^N D_{h/q^\uparrow}$ rescales, approximately, as $1/\sqrt{R^C_{\rm ee}}$
and, as a consequence, $h_1^q(x)$ must rescale as $\sqrt{R^C_{\rm ee}}$.
This rescaling factor
is shown in Fig.~\ref{fig:C-REcc-vs-xi-z02-sc1} for each of the two possible
solutions $\rho_C(\xi_1)$.

This figure shows that in the range of $\xi_1$ considered the rescaling factor
differs from unity by a factor of $\pm 10\%$ at most, that is well inside
the uncertainties of the extraction procedure~\cite{Anselmino:2007fs,
Anselmino:2015sxa}. However, as $z$ increases up to 0.6
the allowed range of $\xi_1$ shrinks to larger values for the FIT07 case, while
for the FIT15 set the rescaling factor $\sqrt{R^C_{\rm ee}}$
decreases down to 0.6 at larger $\xi_1$.

Let us also notice that in this scenario, and within a phenomenological TMD approach,
the possible Collins contribution to SSAs in $p^\uparrow p \to h\,X$,
$p^\uparrow p \to h\,{\rm jet}\,X$ processes should remain approximately unchanged, like
in the SIDIS case, since the transversity distribution and the Collins FF change
simultaneously by an inverse overall factor.

\subsection{\label{sec:col-sc2} Collins Effect, scenario 2}

In this scenario we still require that the two Collins asymmetries for SIDIS
and $e^+e^-$ collisions remain approximately unchanged, Eq.~(\ref{eq:AC-fix}),
imposing this time as a further condition that the transverse momentum dependent
terms of the two Collins asymmetries change in the same way:
\be
\frac{{\cal F}^C_{\rm ee}}{\hat{{\cal F}}^C_{\rm ee}} =
\frac{{\cal F}^C_{\rm DIS}}{\hat{{\cal F}}^C_{\rm DIS}}\, ,
\label{eq:RC-sc2}
\ee
that is, from Eqs.~(\ref{eq:RCDIS}) and (\ref{eq:RCee}):
\be
R^C_{\rm DIS} = R^C_{\rm ee} \>.
\label{eq:RCeqRee}
\ee

\begin{figure}[]
\includegraphics[width=8.truecm,angle=0]{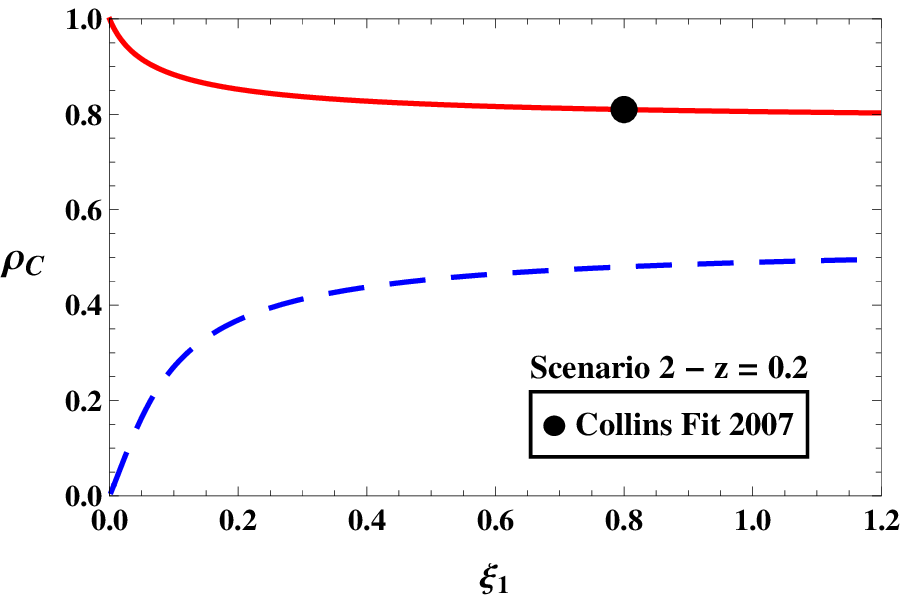}
\includegraphics[width=8.truecm,angle=0]{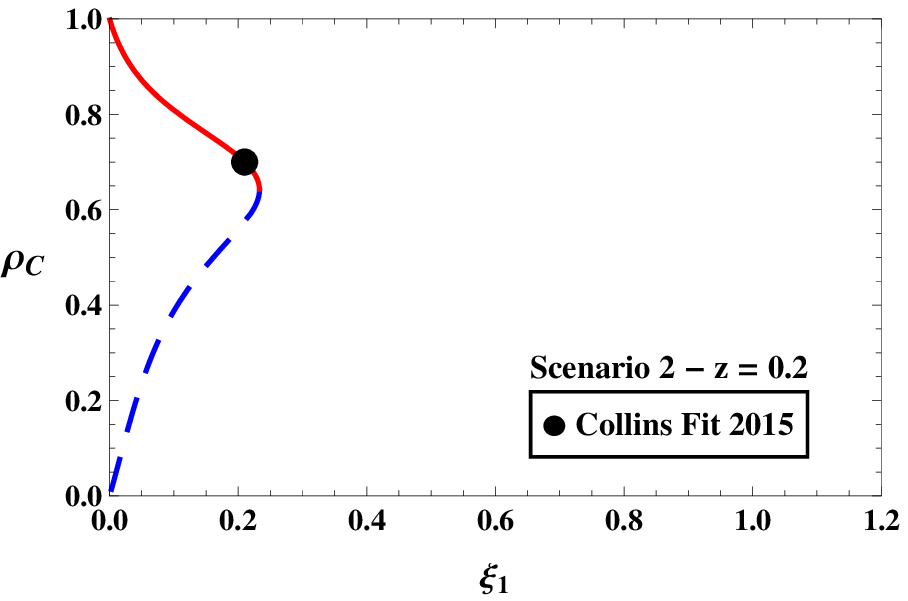}
\caption{The curves in the $(\rho_C,\xi_1$) parameter space show the set of
values of $\rho_C$ and $\xi_1$ which satisfy Eq.~(\ref{eq:RCeqRee}) at $z = 0.2$.
The black dots correspond to the values $\hat{\rho}_C$ and $\hat{\xi}_1$
obtained in the fits of Ref.~\cite{Anselmino:2007fs} (left plot, FIT07) and of
Ref.~\cite{Anselmino:2015sxa} (right plot, FIT15), which describe equally well
the SIDIS and $e^+e^-$ Collins asymmetries. Notice that for each value of $\xi_1$
one finds two possible values of $\rho_C$. Similar results are obtained
by changing $z$ from 0.2 to 0.4 or 0.6.}
\label{fig:C-vs-xi1-z02-sc2}
\end{figure}

At fixed $\hat{\xi}_1$, $\hat{\rho}_C$, $z$, the above constraint translates
into an algebraic cubic equation for $\rho_C(\xi_1)$:
\be
\rho_C^3+\left(\frac{z^2}{\xi_1}-1\right)\,\rho_C^2-\frac{z^2}{\xi_1}\,
\rho_C+\hat{d}(z) = 0\,
\quad\quad {\rm with} \quad\quad
\hat{d}(z) = \hat{\rho}_C(1-\hat{\rho}_C)
\left( \hat{\rho}_C+\frac{z^2}{\hat{\xi}_1} \right)\,.
\label{eq:3th-C-sc2}
\ee
Only 2 out of the 3 solutions are real in the range of $\xi_1$ values of interest.
As an illustration, they are shown in Fig.~\ref{fig:C-vs-xi1-z02-sc2}
for both the FIT07 (left panel) and FIT15 (right panel) parameterisations,
as a function of $\xi_1$ at fixed $z = 0.2$.
Notice that the solutions for the FIT07 case, shown on the left panel,
survive, almost constant, up to values of $\xi_1$ much larger than those shown in the plot.

\begin{figure}[]
\includegraphics[width=8.truecm,angle=0]{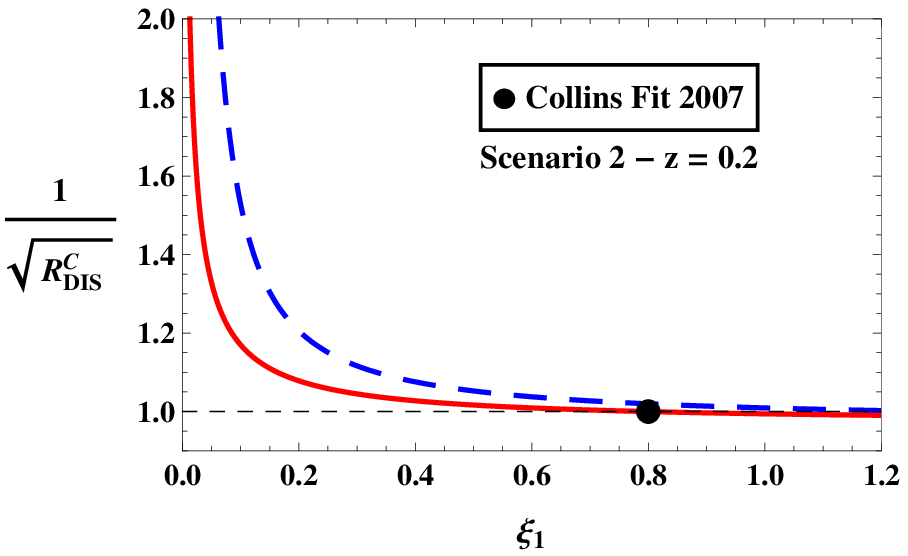}
\includegraphics[width=8.truecm,angle=0]{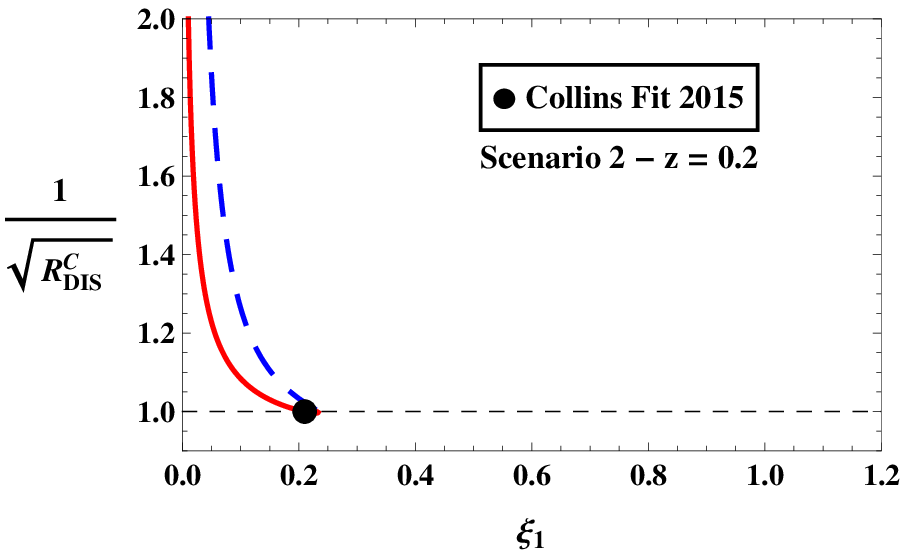}
\caption{The expected rescaling factor $1/\sqrt{R^C_{\rm DIS}}$ for the
collinear transversity distribution $h_1^q(x)$ and the collinear Collins function
$\Delta^N D_{h/q^\uparrow}(z)$ when the parameters $\rho_C,\,\xi_1$ move
away from the reference fit values as in the corresponding plots of
Fig.~\ref{fig:C-vs-xi1-z02-sc2}. In this scenario the total $P_T$-integrated
Collins asymmetries, Eqs.~(\ref{eq:A-DY-AF-int-s}) and (\ref{eq:p0-af-ee-int-s}),
remain unchanged, while ${\cal F}^C_{\rm DIS}$ and ${\cal F}^C_{\rm ee}$
rescale in the same way.}
\label{fig:C-RCee-vs-xi1-z02-sc2}
\end{figure}

The condition (\ref{eq:RCeqRee}) implies
\be
\frac{A^C_{\rm ee}}{\hat{A}^C_{\rm ee}} =
\frac{A^C_{\rm DIS}}{\hat{A}^C_{\rm DIS}} =
\frac{1}{R^C_{\rm DIS}} = \frac{1}{R^C_{\rm ee}} \> ,
\label{eq:RC-sc2}
\ee
which corresponds to a situation in which both the collinear terms
of the transversity distribution, $h_1^q(x)$, and of the Collins fragmentation function,
$\Delta^N D_{h/q^\uparrow}(z)$, are approximately rescaled by the same factor
$1/\sqrt{R^C_{\rm DIS}}$.

For each of the two possible solutions $\rho_C(\xi_1)$ shown in
Fig.~\ref{fig:C-vs-xi1-z02-sc2}, the corresponding rescaling factor
$1/\sqrt{R^C_{\rm DIS}} = 1/\sqrt{R^C_{\rm ee}}$, is shown in
Fig.~\ref{fig:C-RCee-vs-xi1-z02-sc2}, as a function of $\xi_1$ at fixed $z = 0.2$,
for the FIT07 (left panel) and the FIT15 (right panel) cases. For the FIT07 case,
we see that, with the exception of the very small $\xi_1$ region, the rescaling
factor is not far from unity, for both branches of $\rho(\xi_1)$. For the FIT15
case the rescaling factor can be remarkably different for the two solutions
and for one of them can be sizeably larger than unity; however, as we commented
before, the region $\xi_1 \lesssim 0.2$ is unlikely to be a physical one.

As $z$ increases up to 0.6, for the FIT07 case the rescaling factor
for the two solutions differs more and can reach values sensibly different from unity already
for not very small $\xi_1$, of the order $0.4\div0.5$.

\subsection{\label{sec:col-sc3} Collins Effect, scenario 3}

Finally, we consider a scenario based on Eqs.~(\ref{eq:AC-fix}) and
the further constraint that the collinear and the $q_T$-integrated components
of the Collins asymmetry for $e^+e^-$ annihilations remain separately fixed.
According to Eq.~(\ref{eq:RCee}), this corresponds to the condition:
\be
R^C_{\rm ee} = 1 \>,
\label{eq:RCee-sc4}
\ee
that is
\be
\rho_C^3-\rho_C^2+\hat{\rho}_C^2(1-\hat{\rho}_C) = 0\,.
\label{eq:cubic-sc4}
\ee

This equation has the following analytical solutions:
\be
\rho_C = \hat{\rho}_C\,,\quad\quad
\rho_C = \frac{1}{2}\,\left(1-\hat{\rho}_C - \sqrt{1+2\hat{\rho}_C-3\hat{\rho}_C^2}\right)\,,
\quad\quad
\rho_C = \frac{1}{2}\,\left(1-\hat{\rho}_C + \sqrt{1+2\hat{\rho}_C-3\hat{\rho}_C^2}\right)\,.
\label{eq:cubic-sol}
\ee
The second root is always negative in the physical range $0<\hat{\rho}_C<1$,
while the other two take the values:
\bea
&& \rho_C^{(07)1} = \hat{\rho}_C^{(07)} = 0.81 \quad\quad
\rho_C^{(07)2} = 0.50 \label{sol07} \\
&& \rho_C^{(15)1} = \hat{\rho}_C^{(15)} = 0.70 \quad\quad
\rho_C^{(07)2} = 0.63 \label{sol15}
\eea
respectively for the FIT07 and FIT15 cases.

\begin{figure}[]
\includegraphics[width=8.truecm,angle=0]{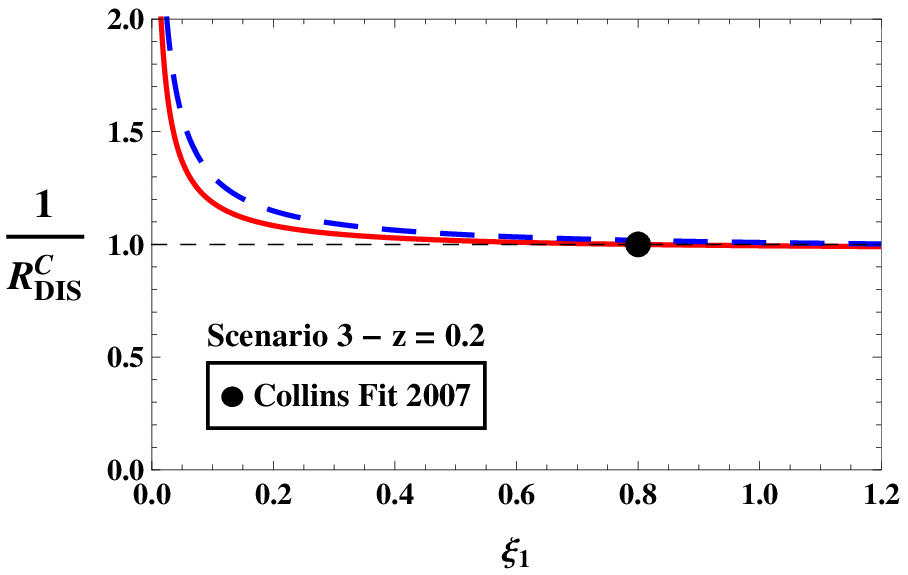}
\includegraphics[width=8.truecm,angle=0]{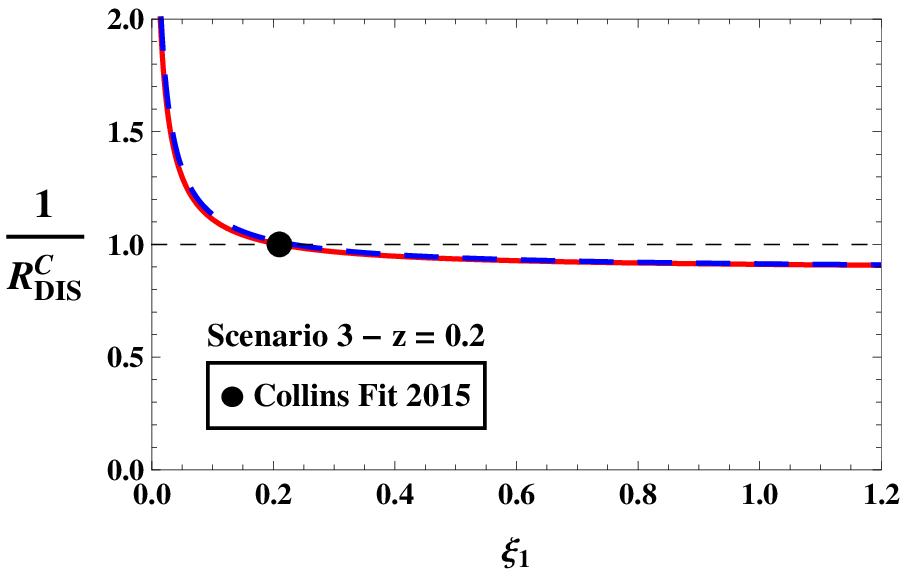}
\caption{The expected rescaling factor $1/{R^C_{\rm DIS}}$ for the
collinear transversity distribution $h_1^q(x)$, as a function of $\xi_1$ at
fixed $z = 0.2$,, when the parameter $\rho_C$ assumes the values given in
Eq.~(\ref{sol07}) (left plot) and in Eq.~(\ref{sol15}) (right plot).
In this scenario the total $P_T$-integrated Collins asymmetries,
Eqs.~(\ref{eq:A-DY-AF-int-s}) and (\ref{eq:p0-af-ee-int-s}), remain unchanged,
as well as the ${\cal F}^C_{\rm ee}$ factor.}
\label{fig:C-RCdis-vs-xi1-sc4}
\end{figure}

{}From Eqs.~(\ref{eq:RCDIS}) and (\ref{eq:RCee}) we have, in this scenario,
\be
A^C_{\rm ee} = \hat{A}^C_{\rm ee} \quad\quad
A^C_{\rm DIS} =  \frac{1}{R^{C}_{\rm DIS}}\,\hat{A}^C_{\rm DIS} \>,
\label{AC-sc-3}
\ee
from which one expects a situation in which the collinear component of the Collins FF,
$\Delta^N D_{h/q^\uparrow}(z)$, remains unchanged, while the transversity
distribution $h_1^q(x)$ changes by a factor $1/R^C_{\rm DIS}$.
The behaviour of this rescaling factor $1/R^C_{\rm DIS}(\rho_C,\xi_1,z)$,
is shown in Fig.~\ref{fig:C-RCdis-vs-xi1-sc4} as a function of $\xi_1$ at fixed
$z=0.2$ for the FIT07 (left panel) and the FIT15 (right panel) cases and, in
each case, for the two allowed $\rho_C$ solutions, Eqs.~(\ref{sol07})
and (\ref{sol15}).

One can see that in both cases the rescaling factor is very similar for
the two possible values of $\rho_C$ and is almost equal to 1, apart from the
small unphysical $\xi_1$ region. As $z$ increases up to 0.6, the difference
between the two solutions is more pronounced for the FIT07 case, and the
rescaling factor differs more sizeably from unity in both cases.

\section{\label{sec:conclusions} Conclusions}

We have investigated to what extent the actual parameterisations of the
most studied TMDs -- the Sivers distribution and the Collins fragmentation
function -- can be fixed by data and what uncertainties could remain.
We feel that such a study is necessary at this stage of the exploration
of the 3D nucleon structure, just before a full implementation of the TMD
evolution is performed and when new amounts
of data are soon expected from COMPASS, JLab 12 and, hopefully in the not
so far future, from the Electron Ion Collider (EIC).

We have done so motivated by the observation that most data originate from
SIDIS processes in which the parton distribution and fragmentation properties
both contribute to build up the final observables, like in Eq.~(\ref{eq:PT-S}),
which clearly shows a strong correlation between $\avk$ and $\avp$. Indeed,
equally good fits of SIDIS asymmetry data could be obtained with rather different values
of these two parameters. On the other hand, other processes, like lepton pair
production in hadronic collisions (DY) or hadron pair production in $e^+e^-$
annihilations, are only sensitive respectively to the TMD parton distributions
or the TMD parton fragmentation functions.

We have assumed a simple scheme, mainly so far adopted, in which the collinear
and transverse degrees of freedom of the TMDs are factorised, with Gaussian
dependences for the transverse momentum dependent components,
Eqs.~(\ref{unp-TMD})--(\ref{Coll-frag}).
We have limited our considerations to the $P_T$ or $q_T$-integrated asymmetries,
Eqs.~(\ref{eq:A-UT-AF-int-s})--(\ref{eq:C-ee}), which have a very simple
structure according to which the Gaussian transverse dependence of the TMDs
results in factors which are functions of the Gaussian widths.
A change in such parameters, like $\avk$ and $\avp$, may affect the extraction
of the collinear part of the TMDs.

We have considered separately the extraction of the Sivers and the Collins
TMDs. The former is related to measured azimuthal asymmetries in polarised SIDIS
and to, so far not yet well known, asymmetries in polarised DY processes.
The latter is related to measured azimuthal asymmetries in polarised SIDIS
and in unpolarised $e^+e^-$ annihilation processes.

We have found that special care must be taken of the uncertainty
in the ratio $\xi_1 = \avp/\avk$ when discussing or adopting the extraction
of the collinear part of the Sivers distribution, $\Delta^N f_{q/\pup}(x)$,
the collinear part of the Collins distribution, $\Delta^N  D_{h/q^\uparrow}(z)$,
or the transversity distribution, $h_1^q(x)$.

In particular, since equally good fits of the Sivers SIDIS asymmetry can be obtained
with considerably different values of $\xi_1$, the extraction of the corresponding
collinear part of the Sivers function, $\Delta^N f_{q/\pup}(x)$, or, equivalently, the prediction of
the Sivers asymmetry in DY processes, may vary by up to a factor 2.
A correct prediction of the Sivers asymmetry in DY processes is of particular importance,
because of the expected sign change of the Sivers function in SIDIS and DY
processes, which remains to be accurately tested.
It is also relevant for the phenomenological study of the TMD evolution of the Sivers distribution.

Concerning the extraction of the collinear component of the Collins function $\Delta^N D_{h/q^\uparrow}(z)$
and the transversity distribution $h_1^q(x)$, from SIDIS and $e^+e^-$ data,
the uncertainty on $\xi_1$ seems to have milder effects. In fact, the rescaling
factors for these functions, when changing the values of the parameters without
altering the quality of the fits, are not far from unity, as shown in
Figs.~\ref{fig:C-REcc-vs-xi-z02-sc1}, \ref{fig:C-RCee-vs-xi1-z02-sc2} and
\ref{fig:C-RCdis-vs-xi1-sc4}. Although our plots cover all mathematically
allowed values of $\xi_1$, down to $\xi_1 = 0$, the physical value of this parameter is
expected to be larger than approximately 0.15~\cite{Bacchetta:2017gcc}.

A precise determination of the parameter $\xi_1 = \avp/\avk$, at least
according to the kinematical configuration of our Gaussian parameterisation,
is of crucial importance for a better knowledge of the Collins, Sivers and
transversity distributions. This parameter enters in the studies of the SIDIS
multiplicities and unpolarised cross section, which then deserve much attention,
both experimentally and phenomenologically. In general, the QCD analysis of the
available data is a formidable task, due to the difficulties in  the correct
implementation of the full theoretical framework and the quality of the experimental results, as
recently pointed out in Ref.~\cite{Boglione:2018dqd}. New important data, helpful in this respect,
might soon be available from JLab 12, COMPASS and future EIC experiments, as well as from Belle, BaBar
and BESIII in the fragmentation sector.

\begin{acknowledgments}
This work was partially supported by the U.S.\
Department of Energy under Contract No.~DE-AC05-06OR23177
and within the TMD Collaboration framework, and by the National
Science Foundation under Contract No.\ PHY-1623454.
\end{acknowledgments}

%


\begin{thebibliography}{10}

\bibitem{Sivers:1989cc}
D.~W. Sivers,
\newblock Phys. Rev. {\bf D41}, 83 (1990).

\bibitem{Sivers:1990fh}
D.~W. Sivers,
\newblock Phys. Rev. {\bf D43}, 261 (1991).

\bibitem{Collins:1992kk}
J.~C. Collins,
\newblock Nucl. Phys. {\bf B396}, 161 (1993).

\bibitem{Airapetian:2009ae}
A.~Airapetian {\em et~al.} (HERMES Collaboration),
\newblock Phys. Rev. Lett. {\bf 103}, 152002 (2009), arXiv:0906.3918.

\bibitem{Adolph:2012sp}
C.~Adolph {\em et~al.} (COMPASS Collaboration),
\newblock Phys. Lett. {\bf B717}, 383 (2012), arXiv:1205.5122.

\bibitem{Allada:2013nsw}
K.~Allada {\em et~al.} (Jefferson Lab Hall A Collaboration),
\newblock Phys. Rev. {\bf C89}, 042201 (2014), arXiv:1311.1866.

\bibitem{Airapetian:2010ds}
A.~Airapetian {\em et~al.} (HERMES Collaboration),
\newblock Phys. Lett. {\bf B693}, 11 (2010), arXiv:1006.4221.

\bibitem{Adolph:2012sn}
C.~Adolph {\em et~al.} (COMPASS Collaboration),
\newblock Phys. Lett. {\bf B717}, 376 (2012), arXiv:1205.5121.

\bibitem{Adolph:2014zba}
C.~Adolph {\em et~al.} (COMPASS Collaboration),
\newblock Phys. Lett. {\bf B744}, 250 (2015), arXiv:1408.4405.

\bibitem{Abe:2005zx}
K.~Abe {\em et~al.} (Belle Collaboration),
\newblock Phys. Rev. Lett. {\bf 96}, 232002 (2006).

\bibitem{Seidl:2008xc}
R.~Seidl {\em et~al.} (Belle Collaboration),
\newblock Phys. Rev. {\bf D78}, 032011 (2008), arXiv:0805.2975.

\bibitem{Aubert:2015hha}
J.~P. Lees {\em et~al.} (BaBar Collaboration),
\newblock Phys. Rev. {\bf D92}, 111101 (2015), arXiv:1506.05864.

\bibitem{TheBABAR:2013yha}
J.~P. Lees {\em et~al.} (BaBar Collaboration),
\newblock Phys. Rev. {\bf D90}, 052003 (2014), arXiv:1309.5278.

\bibitem{Ablikim:2015pta}
M.~Ablikim {\em et~al.} (BESIII Collaboration),
\newblock Phys. Rev. Lett. {\bf 116}, 042001 (2016), arXiv:1507.06824.

\bibitem{Vogelsang:2005cs}
W.~Vogelsang and F.~Yuan,
\newblock Phys. Rev. {\bf D72}, 054028 (2005), arXiv:hep-ph/0507266.

\bibitem{Collins:2005ie}
J.~C. Collins, A.V.~Efremov, K.~Goeke, S.~Menzel, A.~Metz and P.~Schweitzer,
\newblock Phys. Rev. {\bf D73}, 014021 (2006), arXiv:hep-ph/0509076.

\bibitem{Anselmino:2008sga}
M.~Anselmino, M.~Boglione, U.~D'Alesio, A.~Kotzinian, S.~Melis, F.~Murgia, A.~Prokudin and C.~T\"{u}rk,
\newblock Eur. Phys. J. {\bf A39}, 89 (2009), arXiv:0805.2677.

\bibitem{Anselmino:2008jk}
M.~Anselmino, M.~Boglione, U.~D'Alesio, A.~Kotzinian, F.~Murgia, A.~Prokudin and S.~Melis,
\newblock Nucl. Phys. Proc. Suppl. {\bf 191}, 98 (2009), arXiv:0812.4366.

\bibitem{Bacchetta:2011gx}
A.~Bacchetta and M.~Radici,
\newblock Phys. Rev. Lett. {\bf 107}, 212001 (2011), arXiv:1107.5755.

\bibitem{Anselmino:2013vqa}
M.~Anselmino, M.~Boglione, U.~D'Alesio, S.~Melis, F.~Murgia and A.~Prokudin,
\newblock Phys. Rev. {\bf D87}, 094019 (2013), arXiv:1303.3822.

\bibitem{Anselmino:2015sxa}
M.~Anselmino, M.~Boglione, U.~D'Alesio, J.O.~Gonzalez-Hernandez, S.~Melis, F.~Murgia and A.~Prokudin,
\newblock Phys. Rev. {\bf D92}, 114023 (2015), arXiv:1510.05389.

\bibitem{Anselmino:2015fty}
M.~Anselmino, M.~Boglione, U.~D'Alesio, J.O.~Gonzalez-Hernandez, S.~Melis, F.~Murgia and A.~Prokudin,
\newblock Phys. Rev. {\bf D93}, 034025 (2016), arXiv:1512.02252.

\bibitem{Anselmino:2016uie}
M.~Anselmino, M.~Boglione, U.~D'Alesio, F.~Murgia and A.~Prokudin,
\newblock JHEP {\bf 04}, 046 (2017), arXiv:1612.06413.

\bibitem{Anselmino:2007fs}
M.~Anselmino, M.~Boglione, U.~D'Alesio, A.~Kotzinian, F.~Murgia, A.~Prokudin and C.~T\"{u}rk,
\newblock Phys. Rev. {\bf D75}, 054032 (2007), arXiv:hep-ph/0701006.

\bibitem{Bacchetta:2013pqa}
A.~Bacchetta and A.~Prokudin,
\newblock Nucl. Phys. {\bf B875}, 536 (2013), arXiv:1303.2129.

\bibitem{Signori:2013mda}
A.~Signori, A.~Bacchetta, M.~Radici and G.~Schnell,
\newblock JHEP {\bf 11}, 194 (2013), arXiv:1309.3507.

\bibitem{Anselmino:2013lza}
M.~Anselmino, M.~Boglione, J.~O. Gonzalez~Hernandez, S.~Melis and A.~Prokudin,
\newblock JHEP {\bf 04}, 005 (2014), arXiv:1312.6261.

\bibitem{Echevarria:2014xaa}
M.~G. Echevarria, A.~Idilbi, Z.-B. Kang and I.~Vitev,
\newblock Phys. Rev. {\bf D89}, 074013 (2014), arXiv:1401.5078.

\bibitem{DAlesio:2014mrz}
U.~D'Alesio, M.~G. Echevarria, S.~Melis and I.~Scimemi,
\newblock JHEP {\bf 11}, 098 (2014), arXiv:1407.3311.

\bibitem{Kang:2015msa}
Z.-B. Kang, A.~Prokudin, P.~Sun and F.~Yuan,
\newblock Phys. Rev. {\bf D93}, 014009 (2016), arXiv:1505.05589.

\bibitem{Bacchetta:2017gcc}
A.~Bacchetta, F.~Delcarro, C.~Pisano, M.~Radici and A.~Signori,
\newblock JHEP {\bf 06}, 081 (2017), arXiv:1703.10157.

\bibitem{Scimemi:2017etj}
I.~Scimemi and A.~Vladimirov,
\newblock Eur. Phys. J. {\bf C78}, 89 (2018), arXiv:1706.01473.

\bibitem{Collins:2011zzd}
J.~Collins,
\newblock Foundations of Perturbative QCD, Cambridge University Press, Cambridge, UK (2011).

\bibitem{Aybat:2011ge}
S.~M. Aybat, J.~C. Collins, J.-W. Qiu and T.~C. Rogers,
\newblock Phys. Rev. {\bf D85}, 034043 (2012), arXiv:1110.6428.

\bibitem{GarciaEchevarria:2011rb}
M.~G. Echevarria, A.~Idilbi and I.~Scimemi,
\newblock JHEP {\bf 07}, 002 (2012), arXiv:1111.4996.

\bibitem{Echevarria:2012js}
M.~G. Echevarria, A.~Idilbi and I.~Scimemi,
\newblock Phys. Lett. {\bf B726}, 795 (2013), arXiv:1211.1947.

\bibitem{Echevarria:2014rua}
M.~G. Echevarria, A.~Idilbi and I.~Scimemi,
\newblock Phys. Rev. {\bf D90}, 014003 (2014), arXiv:1402.0869.

\bibitem{Rogers:2015sqa}
T.~C. Rogers,
\newblock Eur. Phys. J. {\bf A52}, 153 (2016), arXiv:1509.04766.

\bibitem{Gonzalez-Hernandez:2018ipj}
J.~O. Gonzalez-Hernandez, T.~C. Rogers, N.~Sato and B.~Wang,
\newblock (2018), arXiv:1808.04396.

\bibitem{Anselmino:2011ch}
M.~Anselmino, M.~Boglione, U.~D'Alesio, S.~Melis, F.~Murgia, E.R.~Nocera and A.~Prokudin,
\newblock Phys. Rev. {\bf D83}, 114019 (2011), arXiv:1101.1011.

\bibitem{Bacchetta:2006tn}
A.~Bacchetta, M.~Diehl, K.~Goeke, A.~Metz, P.J.~Mulders and M.~Schlegel,
\newblock JHEP {\bf 02}, 093 (2007), arXiv:hep-ph/0611265.

\bibitem{Anselmino:2009st}
M.~Anselmino, M.~Boglione, U.~D'Alesio, S.~Melis, F.~Murgia and A.~Prokudin,
\newblock Phys. Rev. {\bf D79}, 054010 (2009), arXiv:0901.3078.

\bibitem{Anselmino:2005nn}
M.~Anselmino, M.~Boglione, U.~D'Alesio, A.~Kotzinian, F.~Murgia and A.~Prokudin,
\newblock Phys. Rev. {\bf D71}, 074006 (2005), arXiv:hep-ph/0501196.

\bibitem{Boglione:2017jlh}
M.~Boglione, J.~O. Gonzalez-Hernandez and R.~Taghavi,
\newblock Phys. Lett. {\bf B772}, 78 (2017), arXiv:1704.08882.

\bibitem{Boglione:2018dqd}
M.~Boglione, U.~D'Alesio, C.~Flore and J.O.~Gonzalez-Hernandez,
\newblock JHEP {\bf 07}, 148 (2018), arXiv:1806.10645.

\end{thebibliography}
%
\end{document}